%% file: LensedPrimordial.tex
\documentclass[prd,amsmath,amssymb,floatfix,nofootinbib,superscriptaddress]{revtex4}

\usepackage{graphicx}
\usepackage{bm}
\usepackage{amsmath,amssymb}
\usepackage{epsfig}
\usepackage{color}
\usepackage{natbib}
\usepackage{enumerate}
\usepackage[hypertex]{hyperref}
\usepackage{ifthen}
\usepackage{subfigure}

\input{macros.tex}

\newcommand{\tC}{\tilde{C}}
\newcommand{\Lmin}{{L_{\rm min}}}
\newcommand{\htaunl}{\hat{\tau}_{\rm NL}}
\newcommand{\bl}{{\bf{l}}}
\newcommand{\bx}{{\bf{x}}}
\newcommand{\br}{{\bf{r}}}
\newcommand{\bL}{{\bf{L}}}

\def\eprinttmp@#1arXiv:#2 [#3]#4@{
\ifthenelse{\equal{#3}{x}}{\href{http://arxiv.org/abs/#1}{#1}}{\href{http://arxiv.org/abs/#2}{arXiv:#2} [#3]}}

\providecommand{\eprint}[1]{\eprinttmp@#1arXiv: [x]@}
\newcommand{\adsurl}[1]{\href{#1}{ADS}}
\providecommand{\bibinfo}[2]{\ifthenelse{\equal{#1}{isbn}}{
\href{http://cosmologist.info/ISBN/#2}{#2}}{#2}}

\newcommand{\temp}{T}
\newcommand{\Ltemp}{\tilde{T}}
\newcommand{\LE}{\tilde{E}}
\newcommand{\LB}{\tilde{B}}
\newcommand{\LX}{\tilde{X}}
\newcommand{\LY}{\tilde{Y}}
\newcommand{\LZ}{\tilde{Z}}

\newcommand{\lb}{{\tilde{b}}}

\newcommand{\Cgtwo}{C_{\rm{gl},2}}
\newcommand{\Cgl}{C_{\text{gl}}}
\newcommand{\Cgltwo}{\Cgtwo}

\newcommand{\tf}{\bar{\Theta}}
\newcommand{\Ctot}{C^{\rm{tot}}}

\begin{document}


\title{CMB lensing and primordial squeezed non-Gaussianity}

\author{Ruth Pearson}
\address{Department of Physics \& Astronomy, University of Sussex, Brighton BN1 9QH, UK}

\author{Antony Lewis}
\homepage{http://cosmologist.info}
\address{Department of Physics \& Astronomy, University of Sussex, Brighton BN1 9QH, UK}

\author{Donough Regan}
\address{Department of Physics \& Astronomy, University of Sussex, Brighton BN1 9QH, UK}

\begin{abstract}

Squeezed primordial non-Gaussianity can strongly constrain early-universe physics, but it can only be observed on the CMB after it has been gravitationally lensed. We give a new simple non-perturbative prescription for accurately calculating the effect of lensing on any squeezed primordial bispectrum shape, and test it with simulations.
We give the generalization to polarization bispectra, and discuss the effect of lensing on the trispectrum. We explain why neglecting the lensing smoothing effect does not significantly bias estimators of local primordial non-Gaussianity, even though the change in shape can be $\agt 10\%$. We also show how $\taunl$ trispectrum estimators can be well approximated by much simpler CMB temperature modulation estimators, and hence that there is potentially a $\sim 10$--30$\%$ bias due to very large-scale lensing modes, depending on the range of modulation scales included. Including dipole sky modulations can halve the $\taunl$ error bar if kinematic effects can be subtracted using known properties of the CMB temperature dipole. Lensing effects on the $\gnl$ trispectrum are small compared to the error bar.
In appendices we give the general result for lensing of any primordial bispectrum, and show how any full-sky squeezed bispectrum can be decomposed into orthogonal modes of distinct angular dependence.
\end{abstract}

\date{\today}

\maketitle

\pagenumbering{arabic}

\section{Introduction}

A squeezed bispectrum or trispectrum produced by local primordial non-Gaussianity is observable in the CMB, and if observed would give a powerful way to rule out simple single-field inflation models and strongly constrain general properties of inflation. The local non-Gaussianity produced in the CMB can be thought of as a modulation of small-scale perturbations by large-scale modes, so over a large overdensity there will be more (or less) small-scale power than over an underdensity, depending on the sign of the non-Gaussianity. However if we observe the small-scale modes they will be gravitationally lensed, so in the squeezed limit we expect to see a modulation of the \emph{lensed} small-scale power spectrum due to large-scale modes. This may be important because lensing smooths out acoustic structures, changing the detailed shape of the bispectrum and trispectrum. Previous work Ref.~\cite{Hanson:2009kg} has shown that for the temperature bispectrum due only to local primordial non-Gaussianity, the bias due to this change in shape is very small.
However since the shape is changed, accounting for lensing might be important to correctly identify the form of the non-Gaussianity; for example a different shape which is orthogonal to the unlensed bispectrum will not generally be orthogonal to the lensed bispectrum. In this paper we give new simple approximations for the effect of lensing on the squeezed CMB temperature and polarization bispectra which allow the effects to be calculated easily. We test these approximations against simulations, and quantify the importance of the lensing at different levels of primordial non-Gaussianity.

Previous work has investigated the lensing bispectrum in detail, its potential bias on local non-Gaussianity estimators, and its impact on the variance of the primordial non-Gaussianity estimators~\cite{Smith:2006ud,Serra:2008wc,Hanson:2009kg,Lewis:2011fk}. The bispectrum produced by lensing turns out to be significant, corresponding to a projection of $\fnl\sim 9$ onto the local shape, and should be detectable by Planck. However the effect is easily modelled and subtracted since the detailed shape of the lensing bispectrum is actually very different from the local shape~\cite{Lewis:2011fk}. Here we address the different issue of how lensing affects any other primordial bispectrum that we might want to observe, and assess whether the change due to lensing is important.

We then extend to trispectrum estimators, and estimate the bias on local $\taunl$ and $\gnl$ trispectrum shape due to lensing. We show how this can be calculated easily using simple approximations for the CMB trispectra, and then also discuss the effect of lensing on any primordial local trispectrum. The lensing trispectrum is large, and can be detected at high significance with Planck, but we shall see that its distinctive shape and scale dependence is very different from local primordial non-Gaussianity. We show that although the projection of the lensing shape onto the local primordial shapes gives a signal that is much larger than would be expected in most inflationary models, it is still only a small-fraction of the expected observational error bar.

Aside from the detailed analysis of the lensing effects, we also give a few results of more general interest. In particular we give simple analytic forms for the CMB bispectra and trispectra that are quite accurate in the highly squeezed limit, and show how general squeezed CMB bispectra can be decomposed into modes of distinct angular dependence. We also demonstrate that $\taunl$ can be modelled very accurately as an angular modulation of the small-scale CMB temperature power, and hence can be estimated quickly and nearly optimally using statistical anisotropy estimators. This also clearly shows the importance of the dipole component of the modulation, and how it must be carefully distinguished from the kinematic dipolar modulation.

\section{Lensed squeezed bispectra}

Lensing deflection angles are only a few arcminutes, though coherent on degree scales. As such, lensing only has a large effect on relatively small scales. Local bispectra depend on three wave numbers $\vl_1$, $\vl_2$, $\vl_3$ (we restrict to $l_1\le l_2 \le l_3$ for convenience), and most of the signal is in squeezed triangles with $l_1 \ll l_2,l_3$. It is therefore a good approximation in many cases to take the largest-scale mode to be unlensed: $\Ltemp(\vl_1)\approx \temp(\vl_1)$. This approximation greatly simplifies many calculations with very little loss of accuracy, and also makes a non-perturbative analysis tractable, as shown for the CMB lensing bispectra in Ref.~\cite{Lewis:2011fk}. For the moment we only consider temperature bispectra in the flat-sky approximation, and hence wish to calculate
\be
\la \temp(\vl_1) \Ltemp(\vl_2) \Ltemp(\vl_3) \ra\approx \la \Ltemp(\vl_1) \Ltemp(\vl_2) \Ltemp(\vl_3) \ra = \frac{1}{2\pi} \lb_{l_1 l_2 l_3}\delta(\vl_1+\vl_2+\vl_3),
\ee
where $\lb_{l_1 l_2 l_3}$ is the reduced lensed bispectrum. The approximation was previously called the linear (unlensed) short-leg approximation. In Appendix~\ref{any_shape} we give the general result for any shape, and show that the result for the lensed temperature bispectrum obtained from the linear short-leg approximation is correct to quadratic order in a squeezed expansion.

We can now proceed to calculate the lensed bispectra, following the methods and notation used for calculating the lensed CMB power spectra via lensed correlation functions in Ref.~\cite{Lewis:2006fu}, with
\be
\Ltemp(\vl) = \int \frac{d^2\vx}{2\pi} \temp(\vx+\valpha) e^{-i\vl\cdot \vx}
\label{T_mode}
\ee
where a tilde denotes the lensed field and $\valpha$ is the lensing deflection angle.
Hence in the unlensed short-leg approximation
\be
\la \temp(\vl_1) \Ltemp(\vl_2) \Ltemp(\vl_3) \ra=  \int \frac{d^2\vx_2 }{2\pi}\frac{d^2\vx_3 }{2\pi}\frac{d^2\vl_2' }{2\pi}
\frac{d^2\vl_3' }{2\pi} \la \temp(\vl_1) \temp(\vl_2') \temp(\vl_3') e^{-i\vl_2\cdot \vx_2}e^{-i\vl_3\cdot \vx_3} e^{i\vl_2'\cdot (\vx_2+\valpha_2)} e^{i\vl_3'\cdot (\vx_3+\valpha_3)} \ra.
\ee
The correlation between $\temp(\vl_1)$ and the lensing potentials gives rise to the lensing bispectrum. We are not interested in this term here, and so only keep remaining terms where $\valpha$ can be taken to be uncorrelated to $\temp$.
Hence
\be
\la \temp(\vl_1) \Ltemp(\vl_2) \Ltemp(\vl_3) \ra=  \frac{1}{(2\pi)^2} \int \frac{d^2\vx_2 }{2\pi}\frac{d^2\vx_3 }{2\pi}\frac{d^2\vl_2' }{2\pi}
 b_{l_1 l_2' l_3'}
 e^{i \vx_2\cdot(\vl_2'-\vl_2)}  e^{i \vx_3\cdot(\vl_3'-\vl_3)}
 \la e^{i\vl_2'\cdot \valpha_2} e^{i\vl_3'\cdot \valpha_3} \ra
\ee
where $\vl_3'=-\vl_1-\vl_2'$ and $\valpha_i\equiv \valpha(\vx_i)$. From statistical homogeneity (isotropy on the sky) the expectation value is only a function of $\vr \equiv \vx_2-\vx_3$, so integrating out $\vx_2+\vx_3$ we obtain
\be
\la \temp(\vl_1) \Ltemp(\vl_2) \Ltemp(\vl_3) \ra= \frac{1}{(2\pi)}\delta(\vl_1+\vl_2+\vl_3)  \int \frac{d^2\vr }{2\pi} \frac{d^2\vl_2' }{2\pi}
 b_{l_1 l_2' l_3'}
 e^{i \vr\cdot(\vl_2'-\vl_2)}  \la e^{i\vl_2'\cdot \valpha_2} e^{i\vl_3'\cdot \valpha_3} \ra .
\ee
This is very similar in form to what is required for lensing of the temperature power spectrum~\cite{Seljak:1996ve,Challinor:2005jy,Lewis:2006fu}.
Let's define $\vl'\equiv (\vl_2'-\vl_3')/2 = \vl_2'+\vl_1/2$ and $\vl\equiv (\vl_2-\vl_3)/2 = \vl_2+\vl_1/2$ to encode the wavevectors of the small-scale modes, so that
\be
\vl_2'\cdot \valpha_2 + \vl_3'\cdot \valpha_3 = \vl'\cdot(\valpha_2-\valpha_3) -\frac{\vl_1}{2}\cdot(\valpha_2+\valpha_3).
\ee
Then neglecting non-Gaussianity of the lensing potentials,
\ba
\label{bispeclensedunintegrated}
\lb_{l_1 l_2 l_3} \!&=& \! \! \!\int \frac{d^2\vr }{2\pi} \frac{d^2\vl' }{2\pi}
 b_{l_1 l_2' l_3'}
 e^{i \vr\cdot(\vl'-\vl)}  \exp\left( - \frac{1}{2}\left\la\left[ \vl'\cdot(\valpha_2-\valpha_3) - \frac{\vl_1}{2}\cdot(\valpha_2+\valpha_3)\right]^2\right\ra\right)\\
 &=&
\!\! \!  \int \frac{d^2\vr }{2\pi} \frac{d^2\vl' }{2\pi}
 b_{l_1 l_2' l_3'}
 e^{i \vr\cdot(\vl'-\vl)}  \exp\left( - \frac{1}{2}\left[
 l'{}^2\left(\sigma^2(r) + \cos 2\phi_{l'r} \Cgltwo(r)\right) + \frac{l_1^2}{4}\left(\Cgl(0)+\Cgl(r) - \cos 2\phi_{l_1 r}\Cgltwo(r)\right)
 \right]\right),\nonumber
\ea
where $\sigma^2(r_{32})\equiv\la(\valpha_3-\valpha_2)^2\ra/2$, and $\Cgltwo(r), \Cgl(r)$ are defined as in Ref.~\cite{Lewis:2006fu} (sec. 4.2).

For squeezed shapes the second term in the exponential $\clo(l_1^2 \Cgl(0))$ is very small (same order as things we've already neglected by using the unlensed short-leg approximation) and hence
\ba
\lb_{l_1 l_2 l_3} &\approx&
 \int \frac{d^2\vr }{2\pi} \frac{d^2\vl' }{2\pi}
 b_{l_1 l_2' l_3'} e^{i \vr\cdot(\vl'-\vl)} \exp\left( - \frac{{l'}^2}{2} [ \sigma^2(r) + \cos 2\phi_{l'r} \Cgltwo(r)]\right).
\label{blens_squeezed}
\ea
If $ b_{l_1 l_2' l_3'}$ were a function only of $l_1$ and $|\vl'|$, this could be evaluated trivially using exactly the same form as the result for lensing of the power spectrum.
%
%

More generally we can parameterize the bispectrum in terms of $l_1, l\equiv |\vl_3-\vl_2|/2, \phi_{ll_1}$ instead of $l_1, l_2, l_3$, where $\phi_{ll_1}$  is the angle between $\vl$ and $\vl_1$. We can then expand the angular dependence of the bispectrum as
\be
b_{l_1 l_2 l_3} = \sum_m \bar{b}_{l_1 l}^{m} \, e^{ m i \phi_{ll_1}}
\label{angle_decomp}
\ee
(see Ref.~\cite{Lewis:2011au} for further discussion). From rotational invariance $m$ should be even, and for parity-invariant fields the dependence on $\phi_{l_1 l}$ is only via $|\phi_{l_1 l}|$, so we can equivalently write
\be
b_{l_1 l_2 l_3} = \sum_{m} b_{l_1 l}^{m} \, \cos{ (m\phi_{ll_1})}
\label{angle_decomp2}
\ee
where $m$ is even and $m\ge0$.

Using the expansion of Eq.~\eqref{angle_decomp} in Eq.~\eqref{blens_squeezed}, the angular integrals can then be done giving the lensed bispectrum moments in terms of integrals of modified and unmodified Bessel functions:
\be
\lb_{l_1 l}^{m} \approx \int r dr J_{m}(lr) \int dl'  l' b_{l_1 l'}^{m} e^{-l'{}^2 \sigma^2(r)/2} \sum_n  I_n[l'{}^2 \Cgltwo(r)/2] J_{2n+m}(l'r).
\label{blens_general}
\ee
This shows that lensing, which is on average a statistically isotropic process, does not mix the angular dependence of the squeezed bispectra: the lensed bispectrum $\lb_{l_1 l}^{m}$ depends only the unlensed bispectrum with the same $m$.
For isotropic primordial bispectra in the squeezed limit the angular average $b_{l_1 l}^0$ is expected to dominate over other modes (unless the large-scale modes generate local anisotropy), so to that approximation one would be applying power spectrum lensing to angle-averaged bispectrum slices $b_{l_1 l}^0$ for each $l_1$. Lensing of the $b_{l_1 l}^2$ moments is mathematically identical to power spectrum lensing of the $C^{TE}_l$ power spectrum.

On the flat sky the Fisher correlation between two different bispectra is usually defined (for small signals) by~\cite{Hu:2000ee}
\ba
F(b,b') &=& \frac{1}{2\pi^2} \int l_1 \ud l_1 \int \ud\vl_2^2 \frac{b_{l_1 l_2 l_3} b'_{l_1 l_2 l_3}}{6C_{l_1} C_{l_2} C_{l_3}},
\ea
which is zero if the bispectra are orthogonal. If the bispectra are both squeezed we can expand in terms of angular dependence, and obtain
\ba
F(b,b') =  \frac{1}{\pi} \int l_1 \ud l_1 \int l \ud l  \sum_m \frac{b_{l_1 l}^m b_{l_1 l}^{'m*}}{6C_{l_1} C_{l}^2 }\left(1+\clo(l_1^2/l^2)\right).
\ea
As might be expected bispectrum components with $m\ne m'$ are orthogonal in the squeezed limit. Since lensing does not change the angular dependence, this will remain true after lensing. If the correlation is defined without the power spectra in the denominator (or equivalently for constant white-noise power spectra, or using the bispectrum of whitened fields) this remains true for all triangles.

\subsection{Local non-Gaussianity}

\begin{figure}
\includegraphics[width=10cm]{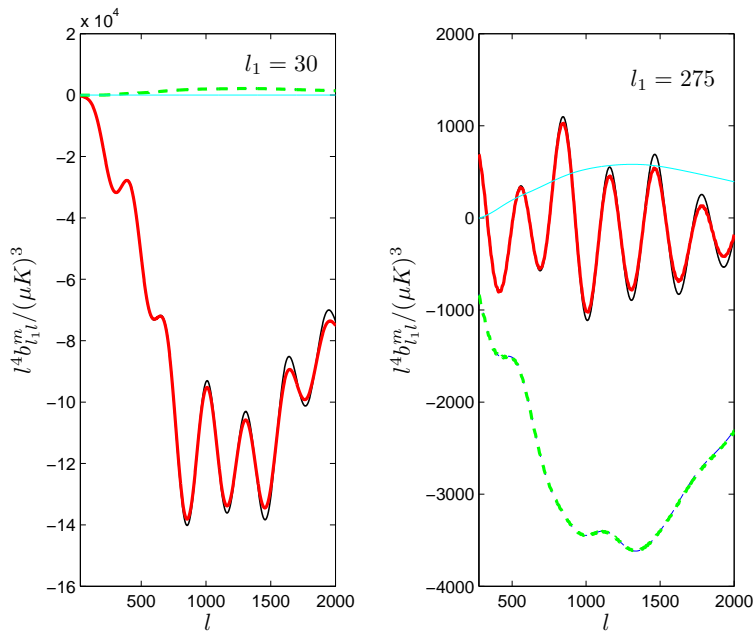}
\caption{The temperature CMB bispectrum from local non-Gaussianity ($\fnl=1$) with $l_1=30$ (left) and $l_1=275$ (right), projected into the isotropic component (solid lines), $m=2$ quadrupole (green) and $m=4$ (cyan) octopole parts. Thick lines are lensed using the approximation described in the text, thin lines show the unlensed bispectrum. The anisotropic components are small on super-horizon
 scales, and on smaller scales are relatively smooth so lensing has little effect. The definition of the angular components on the full sky is given in Appendix~\ref{curved_sky}.}
\label{moment_slices}
\end{figure}

The case of most immediate interest is when the primordial non-Gaussianity is of local (scalar) form, parameterized by $\fnl$. In this form of non-Gaussianity small-scale modes are modulated by the large-scale scalar modes. In the squeezed limit, since the modulation is scalar, the modulation is expected to be isotropic, i.e. $m=0$. Specifically the bispectrum is given by
\bea
b(k_1,k_2,k_3) &=& 2\frac{3}{5}\fnl[P(k_1)P(k_2)+P(k_1)P(k_3) +  P(k_2)P(k_3)]
\nonumber \\&=& 4\frac{3}{5}\fnl P(k_1)P(k)\left[ 1 + \left(\frac{k_1}{k}\right)^2\frac{9+15\cos(2\phi)}{16} +\dots\right],
\label{b_expand}
\eea
where in the second line we expanded in $k_1/k$ for a scale-invariant spectrum, $\phi$ is the angle between the long and the short-scale modes, and $\vk\equiv(\vk_3-\vk_2)/2$. Thus in the squeezed limit there is no $m$
dependence as expected, with leading corrections of $\clo((k_1/k)^2)$ coming from the effect of gradients in the modulation.

Evolution until last scattering will modify this for observations of the CMB. However for $l_1$ corresponding to scales that are super-horizon at recombination ($l_1\ll 100$), the corresponding perturbation can be taken to be constant across the last-scattering surface. The small-scale temperature is then modulated by $1+(6\fnl/5)\zeta_*$
(see more detailed discussion below in Sec.~\ref{mod_taunl}), and hence the local model bispectrum is
\bea
 \la T(\vl_1)T(\vl_2)T(\vl_3)\ra &\approx& C_{l_1}^{T\zeta_*}\left\la \frac{\delta}{\delta\zeta_*(\vl_1)^*}\left( T(\vl_2) T(\vl_3)\right)\right\ra \\
&=&
\frac{1}{2\pi}\delta(\vl_1+\vl_2+\vl_3)\frac{6}{5}\fnl C_{l_1}^{T\zeta_*}( C_{l_2} + C_{l_3}),
\label{local_TTT}
\eea
where $\zeta_*$ is the primordial curvature perturbation at last scattering, $\zeta_*(\vnhat) = \zeta(\vnhat,r_*)$ and $r_*$ is the distance to last-scattering\footnote{Note that $C_l^{T\zeta_*} =\beta_l(r_*)$ with $\beta_l(r)$ defined in Eq.~\eqref{alphabetadef}}. We can see immediately that on the lensed sky, where the lenses are taken to be uncorrelated to the $\zeta_*$, the same result is obtained simply replacing $C_l$ with the lensed $\tilde{C}_l$, and hence it should be no surprise that local bispectrum lensing is well modelled by power spectrum lensing. (Note that this low $l_1$ limit is particularly interesting because non-linear evolution effects are under control analytically and known to be small~\cite{Creminelli:2011sq}.)

When $l_1$ is larger the situation is more complicated since the modulation can no longer be approximated as being constant through last scattering. However the local bispectrum shape is still squeezed, so we can expect that our full result of Eq.~\eqref{blens_general} will accurately describe the effect of lensing where only the lowest angular bispectrum moments are required. Numerically, using the fully-sky analogue of the angular decomposition described in the appendix, we find that the full sky temperature local bispectrum (calculated with {\CAMB}) projected into only its isotropic part is $\sim 96\%$ correlated to the full result ($\lmax=2000$, no noise), and $99\%$ correlated if both the $m=0$ and $m=2$ (isotropic and quadrupolar) moments are retained. The quadrupolar moments are more important for larger (sub-horizon) $l_1\agt 200$. This is because the triangles with a given $l$ have small-scale modes with transfer functions involving $l_2$ and $l_3$ that vary by $\pm l_1/2$ depending on $\phi$, the angle between $\vl_1$ and $\vl$; for $l_1$ comparable to or larger than the separation of the acoustic peaks there is a significant variation in these transfer functions, giving a significant $m\ne 0$ component to the squeezed CMB bispectrum for larger $l_1$. See Fig.~\ref{moment_slices}.

  The unlensed short-leg approximation is expected to be quite accurate since almost all the signal to noise in local non-Gaussianity is at $l_1 \alt 500$ for Planck sensitivity, where lensing effects are still small. We check these approximations numerically below by comparison with full-sky simulations.

\subsection{Polarization}
To consider polarization bispectra in the flat-sky approximation we need the lensed E and B modes given by the polarization analogue of Eq.~\eqref{T_mode}.  Taking the unlensed B modes to be zero, it follows from the definitions given in~\cite{Lewis:2006fu}  that:
\be
\LE(\vl) = \int \frac{d^2\vx}{2\pi}\frac{d^2\vl'}{2\pi} E(\vl') e^{i\vl'\cdot\valpha}e^{i\vx(\vl'-\vl)}\cos{ 2\phi_{l'l}}
\label{E_mode}
\ee
\be
\LB(\vl) = \int \frac{d^2\vx}{2\pi}\frac{d^2\vl'}{2\pi} E(\vl') e^{i\vl'\cdot\valpha}e^{i\vx(\vl'-\vl)}\sin{2\phi_{l'l}}\, ,
\label{B_mode}
\ee
and $\phi_{l'l}$ is the angle between $\vl'$ and $\vl$.
We can then use the lensed E and B modes to calculate the polarization bispectra combinations in the unlensed short-leg approximation as before:
\be
\la X(\vl_1) \LY(\vl_2) \LZ(\vl_3) \ra\approx \la \LX(\vl_1) \LY(\vl_2) \LZ(\vl_3) \ra = \frac{1}{2\pi} \lb_{l_1 l_2 l_3}^{XYZ}\delta(\vl_1+\vl_2+\vl_3),
\ee
where $X$, $Y$ and $Z$ can be any of $T$, $E$ and $B$. For parity invariance fields $b^{TEB}=b^{EEB}=0$.
%

\subsubsection{XTE bispectra}

Following the same steps as for the temperature case, the first combination gives
\be
\la X(\vl_1) \Ltemp(\vl_2) \LE(\vl_3) \ra=  \int \frac{d^2\vx_2 }{2\pi}\frac{d^2\vx_3 }{2\pi}\frac{d^2\vl_2' }{2\pi}
\frac{d^2\vl_3' }{2\pi} \la X(\vl_1) \temp(\vl_2') E(\vl_3') e^{-i\vl_2\cdot \vx_2}e^{i\vl'_2\cdot(\vx_2+ \valpha_2)} e^{i\vl_3'\cdot \valpha_3} e^{i\vx_3\cdot (\vl'_3-\vl_3)} \cos(2\phi_{l_{3}'l_3}) \ra.
\ee
Keeping terms where $\valpha$ is uncorrelated to $T$ and $E$, and writing the expectation value as a function of $\vr \equiv \vx_2-\vx_3$ we have
\be
\la X(\vl_1) \Ltemp(\vl_2) \LE(\vl_3) \ra= \frac{1}{(2\pi)}\delta(\vl_1+\vl_2+\vl_3)  \int \frac{d^2\vr }{2\pi} \frac{d^2\vl_2' }{2\pi}
 b_{l_1 l_2' l_3'}^{XTE}
 \cos(2\phi_{l_{3}'l_3}) e^{i \vr\cdot(\vl_2'-\vl_2)}  \la e^{i\vl_2'\cdot \valpha_2} e^{i\vl_3'\cdot \valpha_3} \ra
\ee
where  $\vl_{3'}=-\vl_1-\vl_2'$. In the squeezed limit, expanding around $\vl_{1}$ gives
$ \cos(2\phi_{l_{3}'l_3})\approx \cos(2\phi_{l'l})+\clo(l_1/l)$.  Hence neglecting non-Gaussianity of lensing potentials and very small terms for squeezed shapes we have
\ba
\lb_{l_1 l_2 l_3}^{XTE} &\approx&
 \int \frac{d^2\vr }{2\pi} \frac{d^2\vl' }{2\pi}
 b_{l_1 l_2' l_3'}^{XTE} \cos(2\phi_{l'l})e^{i \vr\cdot(\vl'-\vl)} \exp\left( - \frac{{l'}^2}{2} [ \sigma^2(r) + \cos 2\phi_{l'r} \Cgltwo(r)]\right).
\label{TTEblens_squeezed}
\ea
Using the expansion of Eq.~\eqref{angle_decomp}, the angular integrals can be done as in the temperature case, giving a result in terms of modified and unmodified Bessel functions:
\be
\begin{split}
\lb_{l_1 l}^{(XTE) m} &\approx \frac{1}{2}\int r dr \int dl'  l' b_{l_1 l'}^{(XTE) m} e^{-l'{}^2 \sigma^2(r)/2} \sum_n  I_n[l'{}^2 \Cgltwo(r)/2] \\& \qquad\qquad\times \left[J_{2n+m+2}(l'r)J_{m+2}(lr)+J_{2n+m-2}(l'r)J_{m-2}(lr)\right].
\label{blens_TTE}
\end{split}
\ee
For the isotropic component $\lb_{l_1 l}^{(XTE) 0}$ the two terms give equal contributions, and the result is then mathematically the same as the result of the lensed $C_l^{TE}$ power spectrum in terms of the unlensed spectrum (see e.g. Ref.~\cite{Lewis:2006fu}). Note however that in general $m$ does not have to be even in the case of the $XTE$ bispectrum, since there is no symmetry between the two small-scale modes. Nonetheless we can expect the main lensing effect to be described in terms of the monopole component, at least for low $l_1$.

\subsubsection{XEE bispectra}
In the case where the two small-scale modes are of the same type, there is a symmetry under interchange of $\vl_2$ and $\vl_3$, and the approximations become valid to $\clo(l_1^2/l^2)$: following the previous argument
\bea
\la X(\vl_1) \LE(\vl_2) \LE(\vl_3) \ra&=&  \int \frac{d^2\vx_2 }{2\pi}\frac{d^2\vx_3 }{2\pi}\frac{d^2\vl_2' }{2\pi}
\frac{d^2\vl_3' }{2\pi} \big\la X(\vl_1) E(\vl_2') E(\vl_3') e^{i\vl_2'\cdot \valpha_2} e^{i\vx_2\cdot (\vl'_2-\vl_2)} e^{i\vl_3'\cdot \valpha_3} e^{i\vx_3\cdot (\vl'_3-\vl_3)}
\cos(2\phi_{l_{3}'l_3})\cos(2\phi_{l_{2}'l_2}) \big\ra \nonumber \\
&=& \frac{1}{(2\pi)}\delta(\vl_1+\vl_2+\vl_3)  \int \frac{d^2\vr }{2\pi} \frac{d^2\vl_2' }{2\pi}
 b_{l_1 l_2' l_3'}^{XEE}
\cos(2\phi_{l_{3}'l_3})\cos(2\phi_{l_{2}'l_2})e^{i \vr\cdot(\vl_2'-\vl_2)}  \la e^{i\vl_2'\cdot \valpha_2} e^{i\vl_3'\cdot \valpha_3} \ra.
\eea
But now in the squeezed limit, expanding in $\vl_{1}$ gives $\cos(2\phi_{l_{3}'l_3})\cos(2\phi_{l_{2}'l_2})=\cos^{2}(2\phi_{l'l})+\clo(l_1^2/l^2)$, and  hence to $\clo(l_1^2/l^2)$
\ba
\lb_{l_1 l_2 l_3}^{XEE} &\approx&
 \int \frac{d^2\vr }{2\pi} \frac{d^2\vl' }{2\pi}
 b_{l_1 l_2' l_3'}^{XEE} \cos^{2}(2\phi_{l'l})e^{i \vr\cdot(\vl'-\vl)} \exp\left( - \frac{{l'}^2}{2} [ \sigma^2(r) + \cos 2\phi_{l'r} \Cgltwo(r)]\right).
\label{TEEblens_squeezed}
\ea
Using the expansion of Eq.~\eqref{angle_decomp}, Eq.~\eqref{TEEblens_squeezed} can then be calculated in terms of integrals of modified and unmodified Bessel functions:
\be
\begin{split}
\lb_{l_1 l}^{(XEE) m} &\approx \frac{1}{4}\int r dr \int dl'  l' b_{l_1 l'}^{(XEE) m} e^{-l'{}^2 \sigma^2(r)/2} \sum_n  I_n[l'{}^2 \Cgltwo(r)/2] \\& \qquad\times \left[2J_{2n+m}(l'r)J_{m}(lr)+J_{2n+m+4}(l'r)J_{m+4}(lr)+J_{2n+m-4}(l'r)J_{m-4}(lr)\right].
\label{blens_TEE}
\end{split}
\ee
Again the isotropic component $\lb_{l_1 l}^{(XEE) 0}$ is given as expected in terms of the unlensed $b_{l_1 l}^{(XEE) 0}$ in exactly the same way as the $C_l^{EE}$ power spectrum is lensed.

\subsubsection{XBB bispectra}
The $\lb^{XBB}$ bispectrum gives a similar result to $\lb^{XEE}$, but we assume it is generated purely by lensing of a primordial $b^{XEE}$ bispectrum:
\ba
\lb_{l_1 l_2 l_3}^{XBB} &\approx&
 \int \frac{d^2\vr }{2\pi} \frac{d^2\vl' }{2\pi}
 b_{l_1 l_2' l_3'}^{XEE} \sin^{2}(2\phi_{l'l})e^{i \vr\cdot(\vl'-\vl)} \exp\left( - \frac{{l'}^2}{2} [ \sigma^2(r) + \cos 2\phi_{l'r} \Cgltwo(r)]\right).
\label{TBBblens_squeezed}
\ea
However since the unlensed $XBB$ signal is expected to be zero for squeezed shapes it is unlikely to be important to model this in the immediate future.

\subsection{Full sky}

So far we have only used the flat-sky approximation. This is very helpful for clarifying the relevant physics and keeping results simple, but in reality the observations that we have are of the full sky. Since squeezed triangles involve large-scale modes, it is therefore important to use a full spherical analysis. In Appendix~\ref{curved_sky} we discuss the equivalent of
Eq.~\eqref{angle_decomp}, and show that as expected bispectra with different angular dependence remain orthogonal in the squeezed limit. A full analysis should then consider how these are lensed.

Focussing on isotropic squeezed bispectra of most interest for primordial non-Gaussianity, we can simply use the understanding from the flat sky that the effect of lensing is just to lens the small-scale power, i.e. we apply the usual curved-sky CMB power spectrum lensing method~\cite{Challinor:2005jy,Lewis:2006fu} to the reduced full-sky bispectrum $b_{l_1 l}^m$, where $l=(l_1+l_2)/2$, separately for each value of $l_1$. The computational cost of this method is then just $N_m\times l_{1,\text{max}}$ times more expensive than lensing the power spectrum, and since the latter operation only costs a fraction of a second on a single CPU this is not problematic if only one or two angular moments ($N_m=1,2$) are required. By contrast the leading perturbative calculation~\cite{Hanson:2009kg} is computationally very challenging, though it has the advantage of also being directly applicable to non-squeezed shapes. In detail we split the unlensed local bispectrum up into $b_{l_1 l}^0$ and $b_{l_1 l}^2$, and a small residual, lens the $m=0$  and $m=2$ bispectra for $l_1<1000$ using a modification of the standard method for lensing the full-sky temperature power spectrum~\cite{Seljak:1996ve,Challinor:2005jy}, and then add back on the small (unlensed) residual part to obtain our estimate of the full lensed bispectrum. Our approximations are not valid for $l_1\gg 500$, and the high $l_1$ angular moments become expensive to calculate, however most of the signal is at lower $l_1$, so we only apply the lensing for $l_1<1000$ and approximate the (very small) contributions from higher $l_1$ as being unlensed. It turns out that the $m=2$ part of the bispectrum is rather smooth and not effected much by lensing, so lensing only the $m=0$ component is usually sufficient (Fig.~\ref{moment_slices}).

Using our simple prescription for lensing the local non-Gaussianity we can then easily calculate various useful results to quantify the importance of lensing. Consider first noise-free temperature data to $\lmax=2000$. We find that the lensed local bispectrum is correlated at the $>0.999$ level with the unlensed bispectrum, with the correction $\delta b$ to the bispectrum due to lensing only biasing estimators based on the unlensed shape by $\sim 0.007\fnl$ (Ref.~\cite{Hanson:2009kg} have previously shown the bias is very small). This confirms that the correction due to lensing is almost orthogonal to the original shape, which should not be surprising since lensing preserves total power (see discussion below). The change in shape due to lensing $\delta b$ is detectable at one sigma for $\fnl\sim 93$. Including polarization data $\delta  b$ would be detectable from lensing of the isotropic component alone for $\fnl\sim 22$, but the bias remains small, $\sim 0.01\fnl$. For Planck noise the current $\fnl$ limit is enough to rule out any chance of detecting the effect of lensing and we confirm the bias is negligible.




\section{Comparison with simulations}

To test our new approximation we compared it with simulations. We generated 480 full-sky Healpix~\cite{Gorski:2004by} maps at $\lmax=2500, n_{\rm side}=2048$ with $\fnl=100$ following the method of Ref.~\cite{Hanson:2009kg}.
The unlensed input power spectra were taken from {\CAMB}~\cite{Lewis:2002ah}. The lensing potential power spectrum was then used to simulate uncorrelated lensing deflection angle maps, and the unlensed maps were then lensed using LensPix~\cite{Lewis:2005tp,Hamimeche:2008ai}.  An estimator for the bispectrum in each full-sky noise-free realization is
\be
\label{ALB}
\hat{B}^{ijk}_{l_{1}l_{2}l_{3}}=\sum_{m_1,m_2,m_3}\begin{pmatrix} l_{1} & l_{2} & l_{3} \\ m_{1} & m_{2} & m_{3} \end{pmatrix}a^{i}_{l_{1}m_{1}}a^{j}_{l_{2}m_{2}}a^{k}_{l_{3}m_{3}},
\ee
where $i$ labels $T$, $E$ or $B$, and we can relate to the reduced bispectrum $b_{l_{1}l_{2}l_{3}}$ defined using:
\be
\label{DHb}
B_{l_{1}l_{2}l_{3}}=\sqrt{\frac{(2l_{1}+1)(2l_{2}+1)(2l_{3}+1)}{4\pi}}\begin{pmatrix} l_{1} & l_{2} & l_{3} \\ 0 & 0 & 0 \end{pmatrix}b_{l_{1}l_{2}l_{3}}.
\ee
To reduce the variance and therefore the number of simulations needed to average over, we follow Ref.~\cite{Hanson:2009kg} by subtracting a term that averages to give zero bispectrum, but removes much of the realization-dependent variance. We also took advantage of symmetries to reduce the number of sums, using:
\begin{equation}
\begin{split}
\label{ourb}
\hat{B}'_{l_{1}l_{2}l_{3}}&= \sum_{m_2=0}^{l_2} \sum_{m_3=-l_3}^{l_3} (2-\delta_{0 m_2}) \threej{l_1}{l_2}{l_3}{m_1}{m_2}{m_3} \Re
\left[  a_{l_{1}m_{1}}a_{l_{2}m_{2}}a_{l_{3}m_{3}} -  \bar{a}_{l_{1}m_{1}}\bar{a}_{l_{2}m_{2}}\bar{a}_{l_{3}m_{3}}  \right]
\end{split}
\end{equation}
where $m_1=-m_2-m_3$, and $a_{lm}$ and $\bar{a}_{lm}$ are the lensed spherical harmonic coefficients from maps generated using the same random seeds but with $\fnl\ne 0$ and $\fnl=0$ respectively. Since we were running multiple simulations, and computational cost is nearly dominated by calculation of the $3j$ symbol, for each set of $\{l,m\}$ we calculated the slice contributions from as many simulations as we could hold in memory.
 To avoid confusion with a lensing-induced bispectrum, the lensing potential is generated with $C_l^{T\psi}=C_l^{E\psi}=0$, so that the subtracted term involving $\bar{a}$ gives zero bispectrum on average.  To test our bispectrum estimation method we compared unlensed simulated bispectra slices to the theory unlensed reduced bispectra generated by {\CAMB}, with good agreement.

 As a first check on polarization bispectra, we used the publicly-available simulated non-Gaussian polarization maps from Ref.~\cite{Elsner:2009md}, which have a maximum multipole of $1024$.

\subsubsection{Simulation results}

For the pure temperature bispectrum $b_{l_{1}l_{2}l_{3}}^{TTT}$ simulations were run up to a maximum multipole of $2000$.  (Although the full-sky Healpix~\cite{Gorski:2004by} maps were $\lmax=2500$, the lensing process requires a few hundred more multipoles than the scale you want to resolve.)  The first acoustic peak of the CMB temperature power spectrum at $l\sim200$ corresponds to the scale which has just had time to maximally compress or expand by the time of recombination.  Therefore a bispectrum with $l_{1}=10$ corresponds to very large super-horizon modulations in the small-scale power; this modulation is roughly constant through the surface of last scattering  (the approximation of Eq.~\ref{local_TTT}).  However a bispectrum with $l_{1}=200$ will give a modulation in small-scale power that varies significantly through the thickness of the last-scattering surface.  We have tested our lensing approximation for both the case of constant and non-constant modulation by simulating bispectra with both $l_{1}=10$ and $l_{1}=200$.

Because of the closure condition, a bispectrum with equal small scale modes ($l_{2}=l_{3}$) looks like an isosceles triangle, where the large scale mode is roughly orthogonal to the short scale modes.  For $l_{3}=l_{1}+l_{2}$ (eg. $b_{10,l,l+10}$), the closure condition demands that the triangle you draw has zero area and all the modes are aligned (parallel).  We test if our lensing approximation holds for both orientations of modes by calculating bispectrum slices for both cases where possible. Since there is a significant quadrupolar $m=2$ part of the bispectrum for $l_1=200$, the slices for the different mode orientations are quite different, and our simulations test that the decomposition of the bispectrum into modes, and lensing just the monopole part, works consistently. For $l_1=10$ the bispectrum is nearly isotropic, so the slices are very similar.

\begin{figure}
\begin{center}
\subfigure{\includegraphics[scale=0.45]{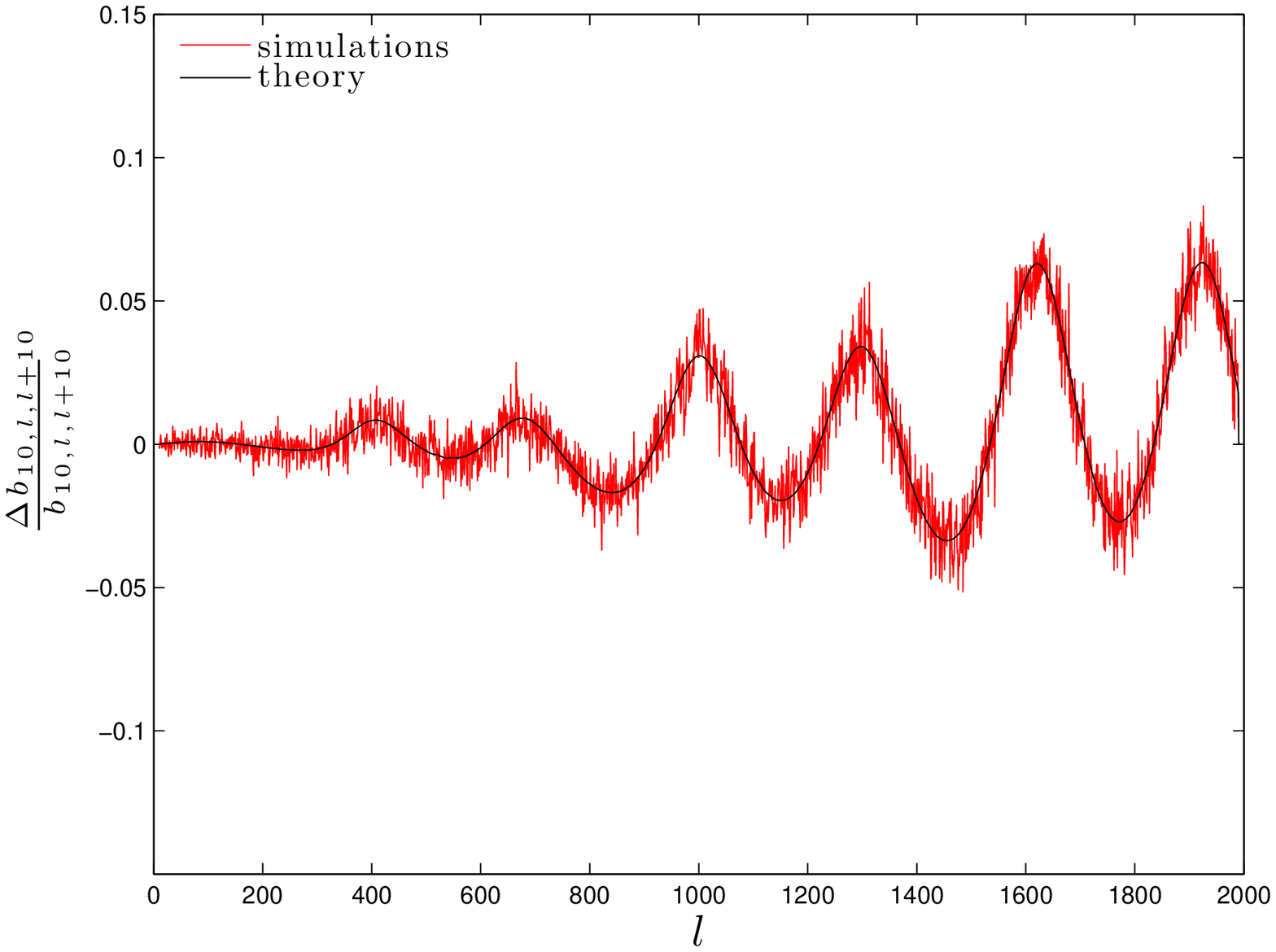}}
\subfigure{\includegraphics[scale=0.45]{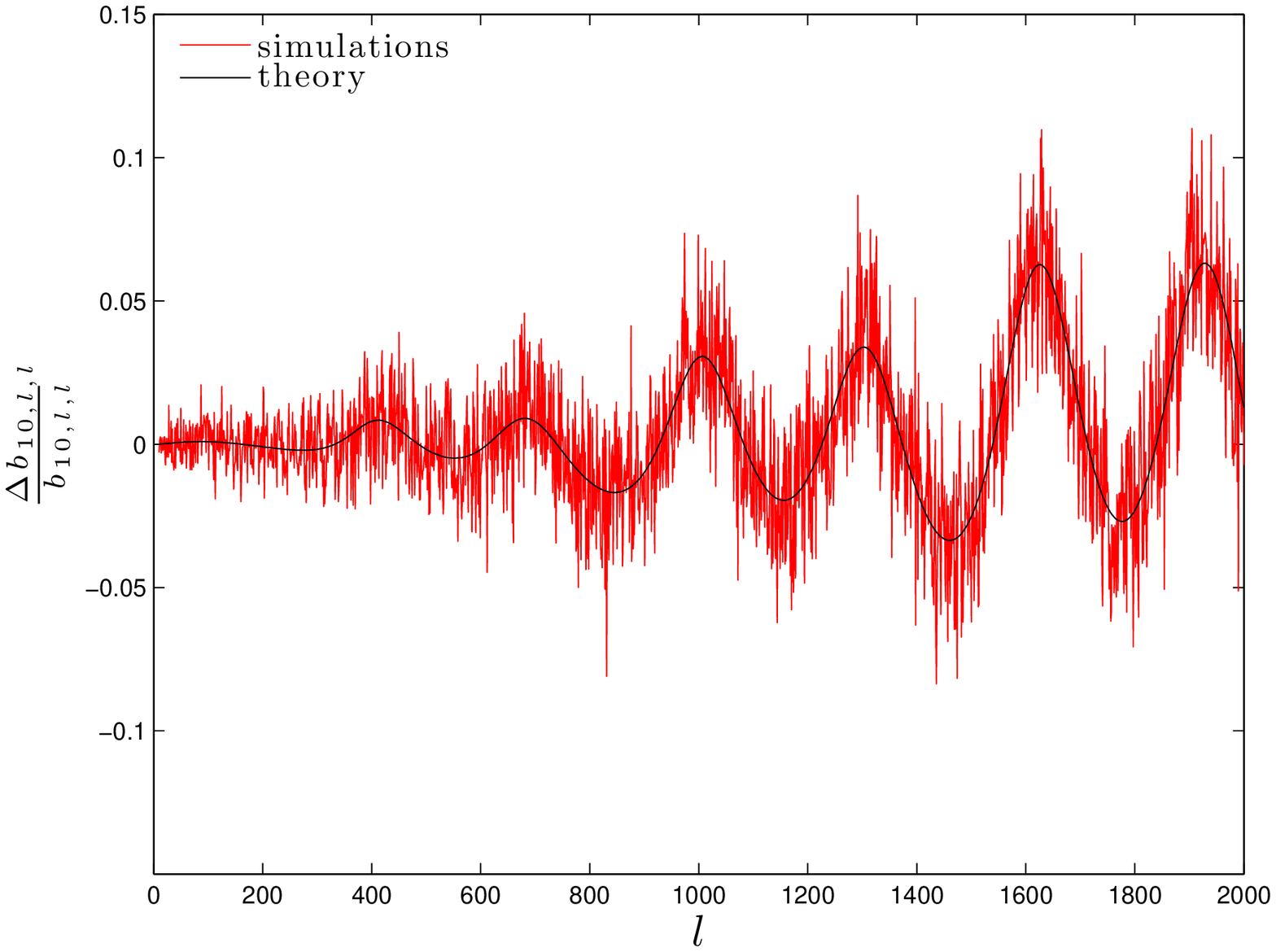}}
\caption{The fractional change due to lensing in the reduced bispectrum slices $b^{TTT}_{10,l,l+10}$ and $b^{TTT}_{10,l,l}$.  The red shows the simulated result averaged over $480$ realizations.  The $b^{TTT}_{10,l,l}$ slice is somewhat noisier than the $b^{TTT}_{10,l,l+10}$ slice (see discussion in text). The black line is the theoretical approximation of this paper, which agrees well with the simulations to within sample variance.}
\label{TTT_10}
\end{center}
\end{figure}

Fig.~\ref{TTT_10} shows the fractional change due to lensing for the reduced bispectrum slices $b^{TTT}_{10,l,l+10}$ and $b^{TTT}_{10,l,l}$, averaged over 480 simulated maps.  In these plots the simulated fractional change has been normalized to remove  variance from the large scale $l_{1}$ modes: since $l_{1}=10$ is a super-horizon mode, its modulation is roughly constant across last scattering and the bispectrum is roughly proportional to $C_{l_{1}}^{T\zeta_{*}}$ (Eq.~\ref{local_TTT}).
Both slices agree very well with the simulations, but the $b^{TTT}_{10,l,l}$ slice has more sampling noise as discussed further below.

\begin{figure}
\begin{center}
\subfigure{\includegraphics[scale=0.45]{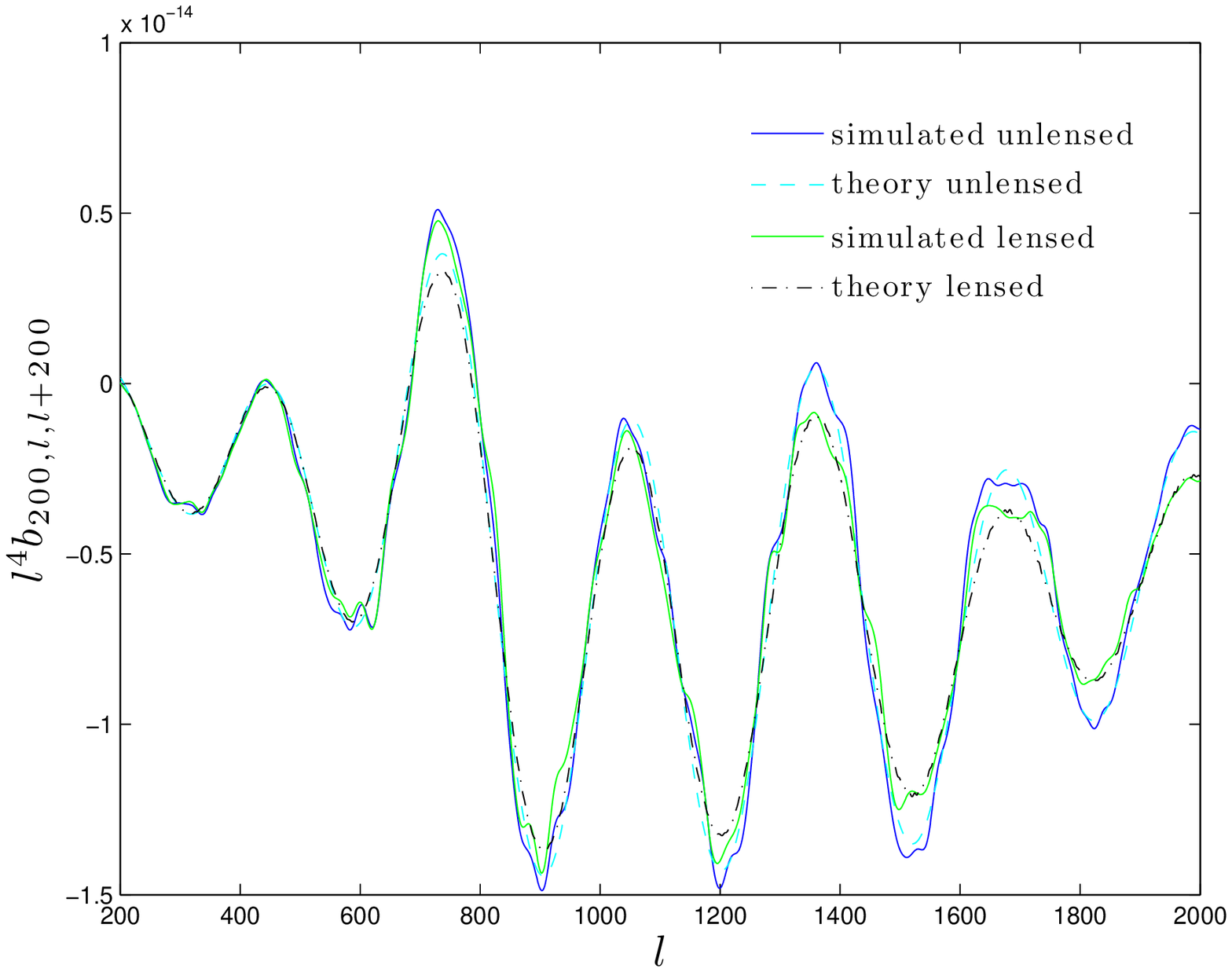}}
\subfigure{\includegraphics[scale=0.45]{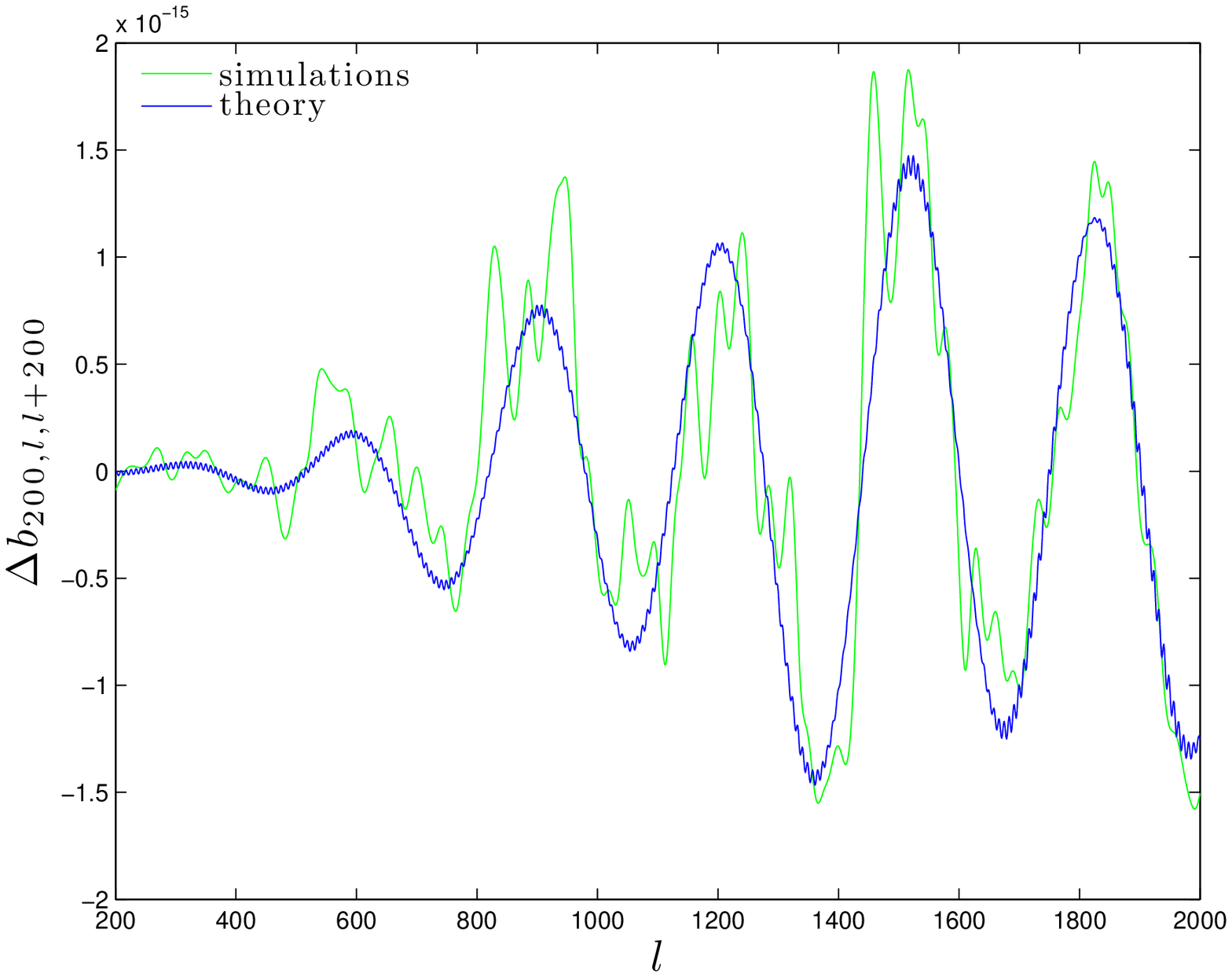}}
\caption{Bispectrum slice $b^{TTT}_{200,l,l+200}$. The left hand panel shows a comparison of the simulated vs theory lensed and unlensed bispectra, averaged over 480 realizations and  smoothed over 10 $l$.  The right panel gives the difference between the lensed and unlensed bispectrum for both simulations and theory, showing good agreement within the sample variance.}
\label{TTT_200}
\end{center}
\end{figure}

 Fig.~\ref{TTT_200} shows the effect of lensing on the $b^{TTT}_{200,l,l+200}$ slice, which also agrees with the theoretical calculation despite the more complicated sub-horizon form of the bispectrum. For larger $l_1$ the bispectrum is no longer isotropic, so $b^{TTT}_{200,l,l+200}$ and $b^{TTT}_{200,l,l}$ differ significantly because of the large $m=2$ quadrupole component.
  Contributions to the signal to noise are roughly equal for every $d \phi$ of angle between the long and the short modes; however since for squeezed shapes $l_3\sim l_2 + l_1\cos(\phi)$, there are more integer $l_3$ per unit angle for orthogonal triangles ($\phi\sim \pi/2$, $l_2\sim l_3$) than parallel ($\phi\sim 0$, $l_2\sim l_3\pm l_1$). The signal to noise per slice is therefore much less for an individual slice with $l_2\sim l_3$ than $l_2\sim l_3\pm l_1$, though the overall contribution to the signal is equally important for the different orientations\footnote{See also footnote in Appendix~\ref{curved_sky}}.
 The simulation sample variance on  $b^{TTT}_{200,l,l}$  was too large to see the lensing effect on a single slice (the fractional effect of lensing is also smaller for triangle shapes to which the smooth $m=2$ bispectrum component contributes significantly; c.f. Fig.~\ref{moment_slices}).

\begin{figure}
\begin{center}
\subfigure{\includegraphics[scale=0.5]{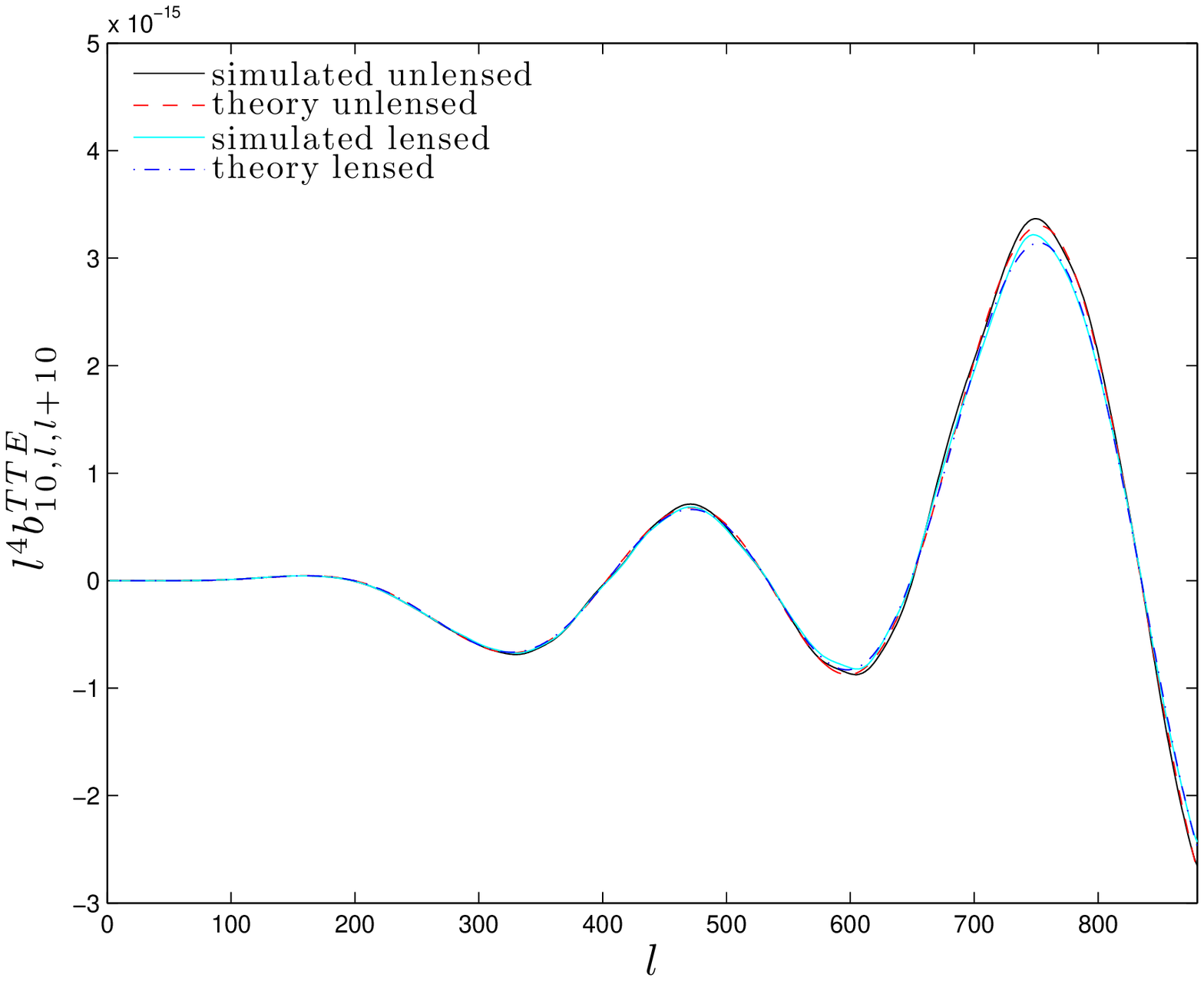}}
\subfigure{\includegraphics[scale=0.5]{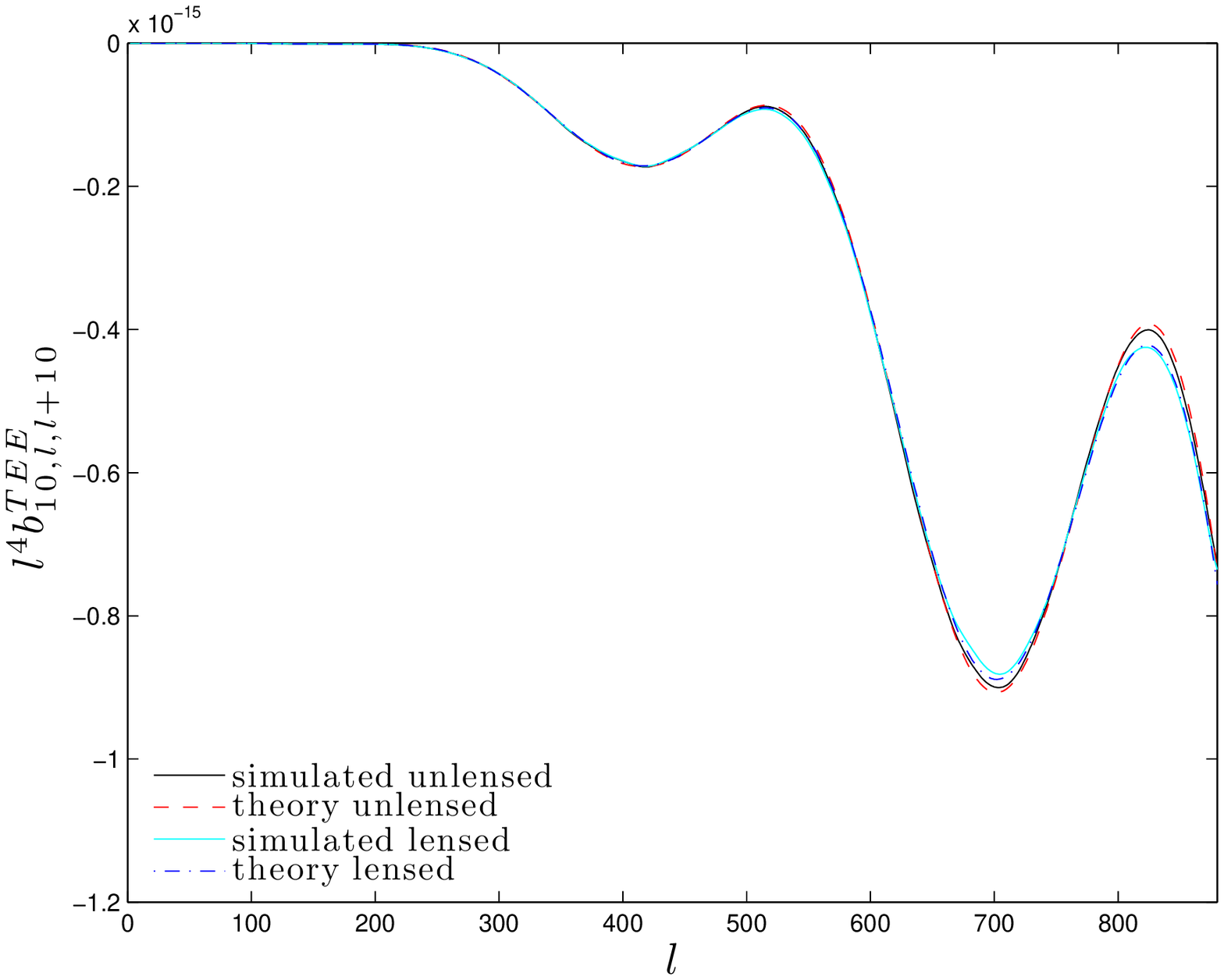}}
\caption{The reduced bispectrum slices $b^{TTE}_{10,l,l+10}$ and $b^{TEE}_{10,l,l+10}$ averaged over 99 realizations and smoothed over 10 $l$.  The black and cyan lines show the simulated and normalized unlensed  and lensed bispectra.  The dashed red line shows the normal theoretical unlensed bispectrum, and the dark blue dot-dashed line shows the lensed approximation of the bispectrum tested in this paper.}
\label{Pol_10_delta_10}
\end{center}
\end{figure}

 Polarization bispectra were estimated up to a maximum multipole of $\sim900$.  Fig.~\ref{Pol_10_delta_10} shows the reduced bispectra for the polarization slices $b^{TTE}_{10,l,l+10}$ and $b^{TEE}_{10,l,l+10}$, which were also normalized to remove large scale variance.  The agreement of the simulations with the theory is seen by eye, as well as the characteristic smoothing of peaks due to lensing.

\subsubsection{Comparison with perturbative result}
\begin{figure}
\includegraphics[width=10cm]{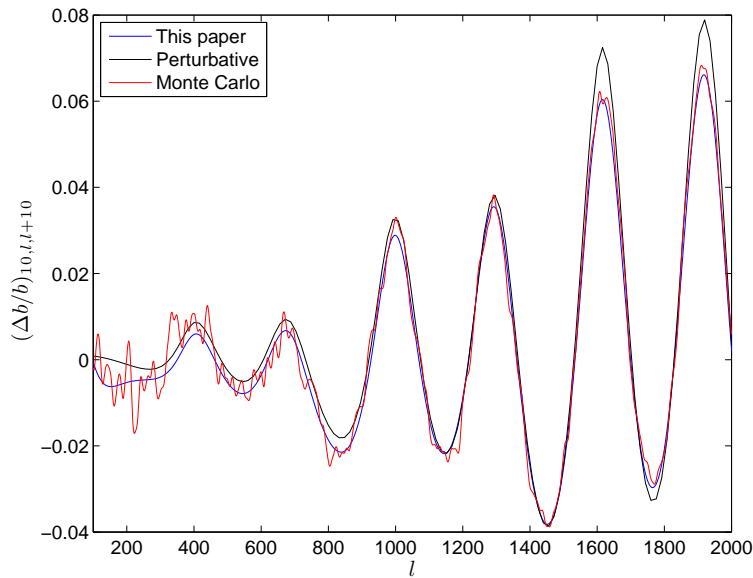}
\caption{The fractional change in the reduced bispectrum slice $b_{10,l,l+10}$ due to lensing. The blue line shows the non-perturbative approximation of this paper, the black line shows the leading-order perturbative result from Ref.~\cite{Hanson:2009kg}. The red lines show the result of 1000  Monte Carlo simulations of Ref.~\cite{Hanson:2009kg} smoothed over $\Delta l=5$. The new approximation only needs to lens the isotropic component of the bispectrum, and then is both significantly more accurate on small scales and faster to compute.
}
\label{perturbative_compare}
\end{figure}

In Fig.~\ref{perturbative_compare} we compare our new approximation to the result from the leading perturbative approximation of Refs.\cite{Cooray:2008xz,Hanson:2009kg} for a squeezed bispectrum slice. The perturbative calculation captures the main effect, but the new approximation is more accurate on small scales; this is essentially exactly the same as the effect of higher-order correction on the power spectrum~\cite{Challinor:2005jy}, since it is the power spectrum that enters the squeezed form of the local bispectrum (Eq.~\eqref{local_TTT}).
Our non-perturbative bispectrum approximation is easier to calculate since it only requires a standard CMB power spectrum lensing calculation for each $l_1$ (only the isotropic part of the bispectrum needs to be lensed to good accuracy). We have only compared with one slice of the full leading perturbative result due to the high numerical cost of calculating the lensed spectra that way.

\section{Trispectrum}
In this section we consider the effect of lensing on local-model trispectrum estimation. The trispectrum depends on the wavevectors $\bl_1,\bl_2,\bl_3,\bl_4$ which define a quadrilateral. The local trispectrum is described by a sum of two shapes - denoted the $g_{NL}$ trispectrum and $\tau_{NL}$ trispectrum respectively, i.e.
\begin{align}
\langle T(\bl_1)T(\bl_2)T(\bl_3)T(\bl_4)\rangle_c=\langle T(\bl_1)T(\bl_2)T(\bl_3)T(\bl_4)\rangle_c^{\gnl}
+\langle T(\bl_1)T(\bl_2)T(\bl_3)T(\bl_4)\rangle_c^{\taunl}.
\end{align}
 In the case of the former, the signal peaks for configurations with two wavenumbers much smaller than the others, e.g. $l_1,l_2\ll l_3,l_4$, while in the latter case the signal peaks when one of the diagonals is small, e.g. $|\bl_1+\bl_2|$,$\ll l_i$. The $\gnl$ shape is independent of the diagonals of the quadrilateral. For simplicity we shall restrict our attention to the temperature trispectra, though in principle better constraints can be obtained making use of polarization information.

 The $\taunl$ shape can be generated with positive sign by scalar quadratic corrections to a Gaussian field; the $\gnl$ shape can be generated with either sign by cubic corrections. If generated by non-linear scalar corrections the squeezed shape dependence is isotropic: $\taunl$ gives a large-scale modulation of all small-scale mode power, independent of the small-scale mode orientation; $\gnl$ looks like a large-scale modulation of a smaller-scale bispectrum~\cite{Lewis:2011fk}. Lensing also generates a large trispectrum~\cite{Hu:2001fa}, which is very blue compared to the local model because the modulation is due to a convergence (and shear) field, rather than a nearly scale-invariant primordial field. The lensing signal is also partly anisotropic due to the shear component; it is primarily a source of confusion for $\taunl$ since lensing does not generate a significant bispectrum except on the largest scales (because the small-scale lenses are uncorrelated to the small-scale anisotropies on the last-scattering surface). The average effect of lensing by many modes does however lead to a smoothing of any primordial trispectrum, similar to the effect we showed in the bispectrum case.

A general trispectrum may be expressed in terms of the five dimensional quantity $p_{l_1 l_2}^{l_3 l_4}(L)$, where
\bea
\langle T(\bl_1) T(\bl_2)T(\bl_3)T(\bl_4)\rangle_c &=&
\frac{1}{(2\pi)^2}\left(
 p_{l_1 l_2}^{l_3 l_4}(|\bl_1+\bl_2|)
+p_{l_1 l_3}^{l_2 l_4}(|\bl_1+\bl_3|)
+p_{l_1 l_4}^{l_2 l_3}(|\bl_1+\bl_4|)\right)\delta(\bl_1+\bl_2+\bl_3+\bl_4).
\label{unlensedtrisp}
\eea
 The $\taunl$ signal peaks in highly squeezed shapes where $L\ll l_1,l_2,l_3,l_4$. If $L=|\vl_1+\vl_2|$ is small, $|\vl_1+\vl_3|$ and $|\vl_1+\vl_4|$ are typically much larger, and hence the trispectrum is well approximated by just the one $p_{l_1 l_2}^{l_3 l_4}(L)$ term, where $\L$ is the wavenumber of the modulation mode.

 The $g_{NL}$ shape is independent of the diagonals $\vL$ of the quadrilaterals and may be expressed in terms of a four dimensional quantity $t_{l_1 l_2 l_3 l_4}$ as
\begin{align}
\langle T(\bl_1)T(\bl_2)T(\bl_3)T(\bl_4)\rangle_c^{g_{NL}}=\frac{1}{(2\pi)^2}t_{l_1 l_2 l_3 l_4}\delta(\bl_1+\bl_2+\bl_3+\bl_4).
\end{align}
This equation may be written in the same form as Eq.~\eqref{unlensedtrisp} by identifying
$t_{l_1 l_2 l_3 l_4}=p_{l_1 l_2}^{l_3 l_4}+p_{l_1 l_3}^{l_2 l_4}+p_{l_1 l_4}^{l_2 l_3}$ with $p_{l_1 l_2}^{ l_3 l_4}=p_{l_1 l_3}^{l_2 l_4}=p_{l_1 l_4}^{l_2 l_3}=t_{l_1 l_2 l_3 l_4}/3$.

\subsection{$\tau_{NL}$ Trispectrum }

The $\taunl$ shape is most easily understood as measuring the power in large-scale modulations of small-scale power; we will show how $\taunl$ can easily be estimated very accurately (and nearly optimally) by making use of modulation estimators. These estimators reconstruct the modulation field on the CMB for each $\vL\equiv \vl_1+\vl_2$, and the trispectrum is then measured by the modulation power spectrum for each value of $\vL$. Using this approach the bias due to lensing and the effect of lensing on the estimators is very easily calculated, and is essentially equivalent to using an accurate approximation for the full trispectrum. For completeness we will then relate to a more direct full calculation. Since the $\taunl$ trispectrum involves nearly full-sky modulation scales, in this section we will do a spherical analysis.

\subsubsection{Modulation estimators for $\taunl$}
\label{mod_taunl}
A trispectrum of $\taunl$ form is generated by a primordial modulation of the fluctuations, where the modulation is not necessarily correlated to the large-scale Gaussian fields. For example if the small-scale Gaussian curvature perturbation $\zeta_0$ is modulated by another field $\phi$ so that
\be
\zeta(\vx) = \zeta_0(\vx)[1+\phi(\vx)]
\ee
where $\phi(\vx)$ is a small large-scale modulating field. The large-scale modes of $\phi$ can be measured by measuring the modulation it induces in the small-scale $\zeta$ power spectrum. If $\phi$ has a nearly scale-invariant spectrum, the nearly-white cosmic variance noise on the reconstruction dominates on small-scales, so only the very largest modes can be reconstructed (e.g. Ref.~\cite{Kogo:2006kh} find almost all the signal in the CMB is at modulation multipoles $l<10$; see Fig.~\ref{tauNLContribs} below). So a constraint on $\phi$ is going to be limited to only very large-scale variations, in which case the scale of the variation is very large compared to the width of the last-scattering surface; i.e. in any particular direction a large-scale modulating field will modulate all perturbations through the last-scattering surface by approximately the same amount. A large-scale power modulation therefore translates directly into a large-scale modulation of the small-scale CMB temperature:
\be
T(\vnhat) \approx T_g(\vnhat)[1 + \phi(\vnhat,r_*)],
\label{Tmodulation}
\ee
where $T_g$ are the usual small-scale Gaussian CMB temperature anisotropies and $r_*$ is the radial distance to the last-scattering surface. This is not quite right for the large-scale temperature due to e.g. the ISW effect, but there are relatively few large-scale modes and they have almost none of the signal to noise in the power modulation, so this error is harmless. But using this rather good approximation we can easily quantify the trispectrum as a function of modulation scale by using the power spectrum of the modulation,
\be
\taunl(L) \equiv \frac{C_L^\phi}{C_L^{\zeta_\star}}.
\label{taunl_L}
\ee
Here were normalized conventionally relative to $C_L^{\zeta_\star}$, the power spectrum of the primordial curvature perturbation at recombination (which we assume is known theoretically, but in practice cannot easily be measured at the relevant scale).
For a scale-invariant primordial spectrum, $\clp_\zeta = A_s$ the angular (\emph{not} CMB!) power spectrum is simply
\be
\frac{L(L+1)C_L^{\zeta_\star}}{2\pi} = A_s.
\ee
Note that $\taunl\sim 500$ corresponds to an $\clo(10^{-3})$ modulation. In Appendix~\eqref{appendix:taunl} we relate this to what you get applying equivalent approximations to the full form of the $\taunl$ trispectrum.

In the simplest local non-Gaussianity model with $\zeta(\vx)=\zeta_0(\vx) + \frac{3}{5}\fnl(\zeta_0(\vx)^2-\la \zeta_0^2\ra)$, we can split $\zeta_0$ into long and short modes $\zeta_l$ and $\zeta_s$, then as far as observations of large-scale modulations are concerned we have
\ba
\zeta&=& \zeta_s(1+ \frac{3}{5}\fnl[2\zeta_l + \zeta_s]) + \zeta_l(1+\frac{3}{5}\fnl\zeta_l)- \frac{3}{5}\fnl\la \zeta_0^2\ra \nonumber\\
&\approx&  \zeta_l + \zeta_s\left(1+ \frac{6\fnl}{5} \zeta_l\right).
\ea
The modulation model for the small-scale modes is then $\phi = \frac{6\fnl}{5} \zeta_l$, and hence $\taunl(L) = (6\fnl/5)^2$ is scale-invariant. Any additional modulation field that is uncorrelated to $\zeta_0$ will not change the bispectrum but will increase $\taunl(L)$, hence in general $\taunl(L) \ge (6\fnl/5)^2$.

Regarding the modulation field as fixed, the small-scale fields being modulated are Gaussian so we can write down a Gaussian likelihood function. We can then try to find the modulation field that maximizes this Gaussian likelihood, giving an estimator for the modulation field. The technical details of how to do this are described in more detail in Ref.~\cite{Hanson:2009gu}, with the result that there is a
quadratic maximum likelihood estimator for the large-scale modulation field $\phi$ given by:
\be
\hat{\phi}= \clf^{-1} \left[ \tilde{\vh}^{\phi}  - \la \tilde{\vh}^{\phi} \ra \right]
\ee
where $\tilde{\vh}$ is a quadratic function of the filtered data that can be calculated quickly in real space:
\be
\tilde{h}^{\phi}_{LM} =
\left.\frac{1}{2}\sum_{l_1 m_2,l_2 m_2} \frac{\delta\la \Ltemp_{l_1 m_1} \Ltemp_{l_2 m_2}^*\ra}{\delta \phi_{LM}^*}\right|_{\phi=0} =
\int d\Omega Y_{LM}^{*} \left[ \sum_{l_1 m_1}\tf_{l_1 m_1} Y_{l_1 m_1} \right] \left[ \sum_{l_2 m_2}\tC_{l_2} \tf_{l_2 m_2} Y_{l_2 m_2} \right].
\ee
Here $\tf = \cov^{-1} \Ltemp$ is the observed CMB sky after inverse-variance filtering (which accounts for sky cuts and inhomogeneous noise),
and $\tC_{l_2}$ is the lensed $C_l$ (since the covariance in the lensed sky involves the lensed power spectra, and we neglect the effect of lensing modes correlated to $\phi_{LM}$). The normalization $\clf$ is given by the Fisher matrix. In the simple full-sky case with isotropic noise this is
\be
\clf_{l m, l' m'} = \delta_{ll'} \delta_{mm'} \sum_{l_1,l_2} \frac{(2l_1+1)(2l_2+1)}{8\pi}
\threejz{l}{l_1}{l_2} ^2 \frac{(\tC_{l_1}+\tC_{l_2})^2}{\Ctot_{l_1} \Ctot_{l_2}},
\ee
and the estimator noise is $N_L = \clf^{-1}_{LL}$. Here $\Ctot_{l_1}$ is the lensed spectrum plus any (isotropic) noise.
For low $L$ and high $\lmax$ the reconstruction noise is very nearly constant (white, because each small patch of sky gives a nearly-uncorrelated but noisy estimate of the small-scale power). We can then define an estimator of the modulation power spectrum
\be
\hat{C}_L^{\phi} = \frac{1}{2L+1} \sum_M | \hat{\phi}_{LM}|^2 - N_L^{(0)},
\label{CLphi}
\ee
where $N_L^{(0)}=\la | \hat{\phi}_{LM}|^2\ra_0$ is a noise bias for zero signal (the same as $N_L$ in the full-sky case). Additional ``$N^{(1)}$'' ($\clo(C^\phi)$) biases from cross-terms are expected to be small, and in any case vanish in the case of zero signal and hence would not give a spurious detection. On the cut sky pseudo-$C_l$ estimators can be used~\cite{Hanson:2009gu,Smidt:2010ra}.

For each value of the modulation scale $L$, Eq.~\eqref{CLphi} gives a separate estimator for $\taunl$. In the ideal case we can combine them by inverse variance weighting with $\var(\hat{C}_L^{\phi}) = 2N_L^2/(2L+1)$, giving the combined estimator
\bea
 \htaunl \equiv \frac{\sum_L \var[\htaunl(L)] ^{-1} \htaunl(L)}{\sum_L \var[\htaunl(L)]^{-1}}
 =  \sigma_{\taunl}^2 \sum_L \frac{(2L+1)C^{\zeta_*}_L}{2N_L} \hat{C}_L^\phi
 \label{taunlest_modulation}
\eea
where $(\sigma_{\taunl}^2)^{-1} = \sum_L (2L+1)/(2N_L[C_L^{\zeta_\star}]^2)$.
Using the approximations $C_L^\zeta \propto (L(L+1))^{-1}$, and the constancy of the white reconstruction noise $N_L$, we have
\be
 \htaunl  \approx  L_{\rm min}^2 \sum_{L=\Lmin}^\infty \frac{2L+1}{L^2(L+1)^2}\frac{\hat{C}_L^\phi}{C_L^{\zeta_\star}}.
\label{taunlest_scaleinv}
\ee
If there is a scale-invariant signal so $C_L^\phi\propto C_L^\zeta$, we see that as expected the contributions fall rapidly $\propto 1/L^3$, as expected when measuring a scale-invariant signal that has large white noise. The total variance is
\be
\var(\hat{\tau}_{NL})\approx \frac{2 L_{\rm min}^2 N_L^2}{(2\pi A_s)^2}.
\label{taunl_var}
\ee
Note that if the $L=1$ dipolar modulation is excluded so that $\Lmin=2$, the $\taunl$ variance becomes four times larger than if the dipole is included. We have $95\%$ of the signal at $L\le 4$, justifying the squeezed approximations used, and hence confirming that they should be good to percent-level accuracy. Assuming the only non-Gaussian signal is $\taunl$, for a noise-free temperature-only experiment to $\lmax =2000$ Eq.~\eqref{taunl_var} gives $\sigma_{\taunl} \sim 150$ (in good agreement with~\cite{Kogo:2006kh} given parameter dependence; note that errors in~\cite{Smidt:2010ra} are a factor of two larger because they exclude the dipole); for Planck $\sigma_{\taunl}\sim 300$. Using numerical values for $C_L^\zeta$ gives somewhat more accurate results for non-scale invariant spectra, with $\sigma_{\taunl} = \left(\sum_L \var[\htaunl(L)] ^{-1}\right)^{-1/2} \approx 134$ for noise-free data to $\lmax=2000$, and $\sigma_{\taunl} \sim 200$--$300$ for Planck depending on assumptions.

The approximation here essentially re-writes the $\taunl$ estimators of Ref.~\cite{Smidt:2010ra} by analytically approximating the radial integrals, resulting in the CMB temperature power anisotropy estimators of Ref.~\cite{Hanson:2009gu}. The approximation consists of taking the recombination visibility as a delta-function compared to the modulation scale of interest, combined with neglect of small cross terms that complicate the relationship between the modulation power spectra and trispectrum (see Ref.~\cite{Kesden:2003cc,Hanson:2010rp} for an extensive discussion in the context of CMB lensing where the latter corrections are more important on small scales). In exchange for these approximations we find estimators that are simple to interpret, fast to evaluate, can incorporate full inverse-variance weighting for optimality with real data, and allow other complicating effects such as lensing to be easily understood and modelled.

The approximate estimator described here accounts for the average effect of small-scale lensing modes in a very simple way, since the only effect is to change the map filter functions so that they involve lensed rather than unlensed power spectra. This is consistent with the expression of the lensed trispectrum given in Appendix~\ref{appendix:taunl}, which only involves power spectra in an equivalent approximation: the large-scale primordial modulation modes cause a modulation in the fully non-linear observed small-scale power. The effect of the modulation on the lensing
potential can be neglected since it only leads to an $\clo(10^{-3})$ correction to the already-small lensing effect.

\subsubsection{Lensing bias}

The above analysis has assumed the only anisotropy was due to primordial modulation. There will also be lensing anisotropy due to large-scale lensing modes (and, for $L=1$, Doppler modulation and angular abberation, as discussed further below). However as discussed in detail in Ref.~\cite{Hanson:2010gu} it is straightforward to account for multiple sources of anisotropy if required, or suboptimally by including lensing in the simulations that are used to subtract noise biases. Here we simply estimate the impact of large-scale lensing modes on a $\taunl$ estimator: i.e. if we neglect lensing, at what level is the simplest $\taunl$ estimator biased?

\begin{figure}
\begin{center}
\includegraphics[width=7cm]{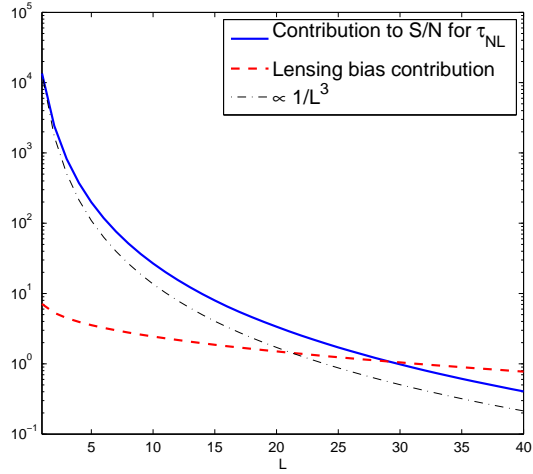}
\caption{Contribution of different modulation scales $L$ to $\taunl$. The signal to noise falls off rapidly with $L$, $\sim 1/L^3$ (solid line). The lensing convergence power spectrum is much bluer than scale invariant, and bias contributions to the $\taunl$ estimator from lensing are nearly flat in $L$; the bias is small if low $L_{\rm max}$ is used, and saturates at around $\la \taunl\ra \sim 100$ if higher $L_{\rm{max}}$ is used. There are also kinematic contributions to $L=1$ that must be removed if $L=1$ is used.}
\label{tauNLContribs}
\end{center}
\end{figure}

Writing the quadratic estimator in harmonic space
\ba
\tilde{h}^{\phi}_{lm}
=
\frac{1}{2}\sum_{l_1,m_1,l_2,m_2}
\sqrt{\frac{(2l+1)(2l_1+1)(2l_2+1)}{4\pi}}
\threejz{l}{l_1}{l_2}
\threej{l}{l_1}{l_2}{m}{m_1}{m_2} (\tC_{l_1}+\tC_{l_2}) \tf_{l_1 m_1}^*
\tf^*_{l_2 m_2}
\label{mod_harmonic}
\ea
we can simply calculate the expectation due to other forms of anisotropy such as lensing. Averaging over small-scale Gaussian lensing and temperature modes $(lm) \neq (l_1 m_1)$ for fixed convergence $\kappa_{l_1 m_1}$ we have~\cite{Okamoto03,Lewis:2011fk}
\begin{multline}
 \sum_{m_2 m_3} \threej{l_1}{l_2}{l_3}{m_1}{m_2}{m_3} \la \Ltemp_{l_2 m_2} \Ltemp_{l_3 m_3}\ra_{(lm)\ne (l_1m_1)} \approx
 \sum_{m_2 m_3} \threej{l_1}{l_2}{l_3}{m_1}{m_2}{m_3}
\left\la \frac{\delta}{\delta \kappa_{l_1 m_1}^*}  \left(\Ltemp_{l_2 m_2} \Ltemp_{l_3 m_3}\right)\right\ra \kappa_{l_1 m_1}^*\\
\approx
\frac{1}{(2l_1+1)}\sqrt{\frac{(2l_1+1)(2l_2+1)(2l_3+1)}{4\pi}} \threejz{l_1}{l_2}{l_3}
\clk_{l_1 l_2 l_3}
\kappa^*_{l_1 m_1},
\end{multline}
where we defined
\be
\clk_{l_1 l_2 l_3} \equiv  (\tC_{l_2}+\tC_{l_3}) + (\tC_{l_2}-\tC_{l_3})\left[\frac{l_2(l_2+1)-l_3(l_3+1)}{l_1(l_1+1)}\right].
\ee
As before all the CMB power spectra are the lensed power spectra. For the simplest full-sky case with isotropic noise $\tf_{lm} = \Ltemp_{lm}/\Ctot_l$,  so the mean field due to lensing is given from Eq.~\eqref{mod_harmonic} and is simply proportional to $\kappa$:
\be
\la \hat{\phi}_{LM}\ra =\kappa_{LM}
\frac{N_L}{2}\sum_{l_1,l_2}
\frac{(2l_1+1)(2l_2+1)}{4\pi}
\threejz{L}{l_1}{l_2}^2
\frac{(\tC_{l_1}+\tC_{l_2})\clk_{L l_1 l_2}}{\Ctot_{l_1} \Ctot_{l_2} }  \equiv \alpha_{L} \kappa_{LM} .
\ee
A positive convergence effectively shifts the CMB power spectrum to the left, decreasing the total power, so $\alpha_L$ is negative, and numerically is typically $\clo(-1)$.
Neglecting some small cross-terms we can then estimate the bias on $C_l^\phi$ due to large-scale lensing modulations by averaging over the lensing modes:
\be
\la \hat{C}_L^\phi\ra \sim \alpha_{L}^2 C_L^\kappa,
\ee
and hence from Eq.~\eqref{taunlest_scaleinv}
\be
 \la \hat{\tau}_{NL}\ra \sim L_{\rm min}^2 \sum_{L=\Lmin}^\infty \frac{2L+1}{L^2(L+1)^2}\frac{\alpha_{L}^2 C_L^\kappa}{C_L^\zeta}
\ee
(or a similar more accurate result using Eq.~\eqref{taunlest_modulation}).

The scaling of the bias contribution with $L$ is shown in Fig.~\ref{tauNLContribs}, in comparison with the contributions to the $\taunl$ signal to noise. The expected bias due to lensing depends on the range of $L$ used. Almost all the $\taunl$ signal is at very low $L$, so a low $L_{\rm max}\alt 10$ includes almost all of the signal to noise. Using this, the Planck lensing bias is $\taunl \sim 17$, or for noise-free data up to $\lmax=2000$ it is $\taunl\sim 40$. If the dipole is not included the bias is relatively larger since the expected modulation is scale-invariant but $\kappa$ is blue: for Planck with $L\ge 2$ the bias is $\taunl\sim 60$. This is still only $\clo(10\%)$ of the size of the error bar, and hence is relatively much less important than for the bispectrum where the lensing signal can give $\fnl\sim 9$~\cite{Hanson:2009kg,Lewis:2011fk}.

However the bias becomes quantitatively more important if a larger $L_{\rm max}$ is used. For example using $L_{\rm max}=40$, $L_{\rm min}=1$  gives $\taunl \sim 80$ for noise-free data, which increases to $\taunl\sim 300$ if $L_{\rm min}=2$: a one-sigma bias. The reason for this is that
although the modulation estimator weighting falls rapidly with $L$, the lensing convergence power is sufficiently blue that its contribution to the bias only falls slowly. This clearly motivates using a lower $L_{\rm max}$ for $\taunl$ analysis, where almost all of the signal is found but the convergence power is small: as we showed above, if $L_{\rm max}\sim 10$ is used the lensing bias is then small enough to be only a small fraction of the error bar.

However as in the case of the bispectrum, the expected bias can also easily be subtracted to obtain an unbiased estimator if required; the simplest method is just to include lensing contributions to $N_L^{(0)}$ in Eq.~\eqref{CLphi} so that the estimator has zero expectation over realizations of lensed skies. If $\taunl$ is non-zero, then there will also be cross-terms from universe-sized modes generating correlated low multipole signals in both the lensing potential and the primordial modulation.

The Doppler dipole due to the earth's motion is $\sim 10^{-3}$, and induces aberration and CMB temperature modulation at this level~\cite{Challinor:2002zh,Kosowsky:2010jm,Amendola:2010ty}. If data are used to constrain $\taunl$ using the dipole modulation (which shrinks the error bar by a factor of two relative to starting at $L=2$), the dipole-induced signal must be subtracted since its modulation reconstruction has signal-to-noise larger than one at Planck resolution. However we know the direction of the kinematic dipole in the CMB temperature, and hence both the magnitude and direction of the modulation that is expected; it can therefore be directly subtracted from the reconstructed dipole modulation field, and as long as this can be done to $\clo(10\%)$ accuracy it should not be an obstacle to also detecting a primordial $\taunl$ that is detectable in the absence of kinematic effects. If the dipole is not used, the dominant primordial signal is then in the quadrupole, which may be susceptible to observational artefacts due to the quadrupolar dependence of Planck scan strategy, for example due to gain variations. Even in the absence of unexpectedly large primordial signals, tests for modulations can be a useful probe of instrumental systematics.

\subsubsection{Lensing bias from simulations and full $\taunl$ estimators}
 To check consistency of the various approximations, we implement the trispectrum estimator as described in Ref.~\cite{Regan:2010cn} and estimate the lensing bias from simulations. The estimator was tested using $300$ Gaussian maps, which were then lensed using LensPix as before.

The averaged full-sky estimator with isotropic noise is given by
\begin{align}\label{Estimator}
\langle\mathcal{E}\rangle=\frac{1}{8F}\sum_{l_i m_i}\sum_{L M}\frac{p^{l_1 l_2}_{l_3 l_4}(L)(-1)^M \mathcal{G}^{l_1 l_2 L}_{m_1 m_2 -M}\mathcal{G}^{l_3 l_4 L}_{m_3 m_4 M} }{\Ctot_{l_1}\Ctot_{l_2}\Ctot_{l_3}\Ctot_{l_4}}\Bigg[&\langle a_{l_1 m_1}a_{l_2 m_2}a_{l_3 m_3}a_{l_4 m_4}\rangle-\langle a_{l_1 m_1}a_{l_2 m_2}\rangle\langle a_{l_3 m_3}a_{l_4 m_4}\rangle\nonumber\\
&-\langle a_{l_1 m_1}a_{l_4 m_4}\rangle\langle a_{l_2 m_2}a_{l_3 m_3}\rangle-\langle a_{l_1 m_1}a_{l_3 m_3}\rangle\langle a_{l_2 m_2}a_{l_4 m_4}\rangle\Bigg],
\end{align}
where the angular brackets $\langle \dots \rangle$ refer to averaging over the simulations and the Gaunt integral is given by $\mathcal{G}^{l_1 l_2 l_3}_{m_1 m_2 m_3}=\int d\hat{\bf{n}}Y_{l_1 m_1}Y_{l_2 m_2}Y_{l_3 m_3 }$. The denominator is given by the Fisher matrix,
\begin{align}
F=\frac{1}{8}\sum_{l_i L}\frac{P^{l_1 l_2}_{l_3 l_4}(L) }{\Ctot_{l_1}\Ctot_{l_2}\Ctot_{l_3}\Ctot_{l_4}}\Big(\frac{P^{l_1 l_2}_{l_3 l_4}(L)}{2L+1}+\sum_{L'}&(-1)^{l_2+l_3}\left\{ \begin{array}{ccc}
l_1 & l_2 & L \\
l_4 & l_3 & L' \end{array}\right\} P^{l_1 l_3}_{l_2 l_4}(L')+\sum_{L'}(-1)^{L+L'}\left\{ \begin{array}{ccc}
l_1 & l_2 & L \\
l_3 & l_4 & L' \end{array}\right\} P^{l_1 l_4}_{l_3 l_2}(L')\Big),
\end{align}
where $\{ \dots \}$ refer to the Wigner $6j$ symbols, and $P^{l_1 l_2}_{l_3 l_4}(L) =p^{l_1 l_2}_{l_3 l_4}(L)h_{l_1 l_2 L}h_{l_3 l_4 L} $, with $h_{l_1 l_2 L}\equiv \sqrt{\frac{(2l_1+1)(2l_2+1)(2L+1)}{4\pi}}
\threejz{l_1}{l_2}{L}$.
In the case of the local model $\taunl$ trispectrum (given by \eqref{eq:taunltrisp}), the squeezed shape dependence means that only
$P^{l_1 l_2}_{l_3 l_4}(L)$ contributes significantly, and $F$ may be efficiently and accurately calculated using \cite{Kogo:2006kh},
\begin{align}
F=\frac{1}{8}\sum_{l_i L}\frac{p^{l_1 l_2}_{l_3 l_4}(L)^2 h_{l_1 l_2 L}^2 h_{l_3 l_4 L}^2 }{\Ctot_{l_1}\Ctot_{l_2}\Ctot_{l_3}\Ctot_{l_4}}.
\end{align}
The optimal signal to noise of the trispectrum is given by $F$, and the corresponding variance of $\taunl$ is given by its inverse. For $\lmax=2000$ the error bar is given by $\sigma(\taunl)=129$, where the sum over $L$ has been calculated for $1\leq L \leq 100$. We checked that the sum has converged to within $\sim2\%$ of its final value by $L=5$ and to within $\sim 0.5\%$ by $L=10$. Furthermore, the replacement of $F_L(r_1,r_2)$ by $F_L(r_*,r_*)=C_L^{\zeta_*}$, as described in Appendix~\ref{appendix:taunl} results in little loss of accuracy, with the variance in $\taunl$ agreeing with the previous estimate to within $\mathcal{O}(1\%)$. Calculating the full trispectrum of Eq.~\eqref{eq:taunltrisp} however involves doing radial integrals, and the numerical error involved can be larger than the theoretical approximations made in the previous sections. The direct calculation of $\sigma(\taunl)=129$ agrees quite well with the previous modulation-estimator value of  $\sigma(\taunl)=134$, with a small portion of the difference being made up by $\clo(1\%)$ theoretical approximations, and few percent numerical errors.

Using $F_L(r_1,r_2) \sim F_L(r_*,r_*)$, the estimator for the amplitude of the local $\taunl$ model reads
\begin{align}
\hat{\tau}_{NL} =&\frac{1}{2F}\sum_{L}F_L(r_*,r_*) \Big[\sum_M |(A*B)_{L M}|^2
- \sum_{l_1 l_2} \frac{h_{l_1 l_2 L}^2 }{2\Ctot_{l_1}\Ctot_{l_2}}AB(l_1,l_2)^2\Big],
\label{taunlest}
\end{align}
where
\begin{align}
(A*B)_{L M}&=\int d\hat{\bf{n}}Y_{L M}^*(\hat{\bf{n}})\int dr r^2 A(r,\hat{\bf{n}})B(r,\hat{\bf{n}}),\\
A(r,\hat{\bf{n}})&=\sum_{l_1 m_1}\frac{\alpha_{l_1}(r) a_{l_1 m_1}Y_{l_1 m_1}(\hat{\bf{n}})}{\Ctot_{l_1}},\\
B(r,\hat{\bf{n}})&=\sum_{l_1 m_1}\frac{\beta_{l_1}(r) a_{l_1 m_1}Y_{l_1 m_1}(\hat{\bf{n}})}{\Ctot_{l_1}},\\
AB(l_1,l_2)&=\int dr r^2\left( \alpha_{l_1}(r)\beta_{l_2}(r)+\alpha_{l_2}(r)\beta_{l_1}(r)\right),
\end{align}
and where $\Ctot_{l_1}=C_{l_1}$ for unlensed Gaussian maps, but is replaced by the lensed angular power spectrum for the lensed Gaussian maps (plus noise if applicable). Using the further accurate approximation of Eq.~\eqref{alphabetaint}, Eq.~\eqref{taunlest} is consistent with the previous modulation estimator of Eq.~\eqref{taunlest_modulation} for the low $L$ of interest. In fact we have verified that use of this approximation in the estimator recovers the same variance and bias to $\mathcal{O}(1-2\%)$.

The variance and mean of the estimator has been calculated over 300 Gaussian simulations giving a sampling bias of $3$, and a one sigma error of $130$. This error bar agrees quite well with the expected variance of $129$.

Applying the estimator up to $L_{\rm{max}}=10$ to the lensed maps reveals a bias of $49$ and a slightly enlarged variance of $141$. Hence, application of the equation \eqref{Estimator} verifies that large-scale lensing modes may introduce a small bias on $\taunl$ estimators of $\mathcal{O}(30\%)$ the size of the error bar for noise-free data out to $\lmax=2000$. Extending the summation to $L_{\rm{max}}=100$ increases the bias to $107$, confirming the expectation that the bias would become quantitatively larger due to the blue spectrum of the lensing convergence.
A realistic analysis on the lensed sky can restrict to a much lower value of $L_{\rm{max}}$ to keep the bias small, with very little loss of $\taunl$ signal.

We conclude that the modulation estimator is fully consistent with a more brute-force $\taunl$ estimator, and in fact that the rather accurate theoretical approximations $(\alt 1\%)$ involved are likely to be more accurate than numerical errors involved in the full estimator (which involves performing numerical radial integrals). Both are accurate to within a small fraction of the error bar, but the modulation estimator is significantly faster to calculate and makes it much simpler to calculate complications such as lensing bias.

\subsubsection{Lensing of the $\taunl$ trispectrum}
We have shown how to easily account for the main effects of lensing when estimating $\taunl$, and quantified the bias due to overlap between the $\taunl$ and lensing trispectrum shapes. Here we briefly consider how this is consistent with the direct result for lensing of a primordial $\taunl$ trispectrum.

The lensed $\tau_{NL}$ trispectrum is defined by
\begin{align}
\langle \Ltemp(\bl_1)\Ltemp(\bl_2)\Ltemp(\bl_3)\Ltemp(\bl_4)\rangle_c=\frac{1}{(2\pi)^2}\left(\tilde{p}_{l_1 l_2}^{l_3 l_4}(|\bl_1+\bl_2|)+
\tilde{p}_{l_1 l_3}^{l_2 l_4}(|\bl_1+\bl_3|)
+\tilde{p}_{l_1 l_4}^{l_2 l_3}(|\bl_1+\bl_4|)\right)\delta(\bl_1+\bl_2+\bl_3+\bl_4).
\end{align}
In terms of the unlensed trispectrum we have
\begin{align}
\langle \Ltemp(\bl_1)\Ltemp(\bl_2)\Ltemp(\bl_3)\Ltemp(\bl_4)\rangle_c^{}=&\int \Pi_{i=1}^4 \left( \frac{d^2 \bx_i d^2 \bl_i'}{(2\pi)^2}\right) \delta(\sum_i\bl_i')\frac{1}{(2\pi)^2}\left[{p}_{l_1' l_2'}^{l_3' l_4'}(|\bl_1'+\bl_2'|)+(2\,\,{\rm{perms}})\right]\nonumber\\
&\times e^{i(\bl_1'-\bl_1).\bx_1}e^{i(\bl_2'-\bl_2).\bx_2}e^{i(\bl_3'-\bl_3).\bx_3}e^{i(\bl_4'-\bl_4).\bx_4} \langle e^{i\bl_1'.\valpha_1} e^{i\bl_2'.\valpha_2} e^{i\bl_3'.\valpha_3} e^{i\bl_4'.\valpha_4}\rangle.
\end{align}
Using statistical homogeneity we may assume that the expectation value is independent of $\sum_i \bx_i$. Defining $\vl_{43} \equiv (\vl_4-\vl_3)/2$, $\vl_{21}\equiv (\vl_2-\vl_1)/2$, $\vL\equiv \vl_1+\vl_2$, $\vr \equiv (\vx_1+\vx_2-\vx_3-\vx_4)/2$ this allows us to write
\begin{multline}
\label{taunlLens}
\tilde{p}_{l_1 l_2}^{l_3 l_4}(L)=\int \frac{d^2 \br_{12}d^2 \br_{34}d^2 \br d^2\bl_{21}' d^2\bl_{43}' d^2\vL' }{(2\pi)^6} p_{l_1' l_2'}^{l_3' l_4'}( L')
e^{i\br_{21}.(\bl_{21}'-\bl_{21})}e^{i\br_{43}.(\bl_{43}'-\bl_{43})}e^{i\br.(\bL'-\bL)} \\
\times \langle e^{i\bl_{21}'.\valpha_{21}}e^{i\bl_{43}'.\valpha_{43}}e^{i \half\vL' \cdot (\valpha_1+\valpha_2-\valpha_3-\valpha_4)}\rangle,
\end{multline}
where we assumed a squeezed shape, $L',L\ll l_i,l_i'$ and so only kept $\tilde{p}_{l_1 l_2}^{l_3 l_4}(L)$. Also for squeezed shapes we can neglect the $\vL'$ term in the exponential (as for the bispectrum), and we have
\begin{align}
 \langle e^{i\bl_{21}'.\valpha_{21}}e^{i\bl_{43}'.\valpha_{43}}\rangle
=\exp\Big(-\frac{1}{2}\langle (\bl_{21}'.\valpha_{21})^2\rangle\Big) \exp\Big(-\frac{1}{2}\langle(\bl_{43}'.\valpha_{43})^2\rangle\Big)\exp\Big(-\langle (\bl_{21}'.\valpha_{21})(\bl_{43}'.\valpha_{43})\rangle \Big).
\label{expapprox}
\end{align}
The first two terms are exactly the terms that appear in bispectrum lensing (c.f. Eq.~\eqref{bispeclensedunintegrated}), and hence are well approximated by power spectrum lensing of the small-scale power at wavenumbers $l_{21}$ and $\l_{43}$. The last term is an additional effect coming from the fact that the lenses acting on different pairs of small-scale modes are correlated (all small-scale modes see the same lensing potential realization).
It depends on $\vr$, and hence in principle gives rise to couplings between $\vL\ne \vL'$; however we are interested in very low $L$, corresponding to a scale much larger than the arcminute lensing deflections, so we can expect $\vL\approx \vL'$ to a good approximation for all trispectrum components of interest. This is consistent because the difference in deflection angles at $\vx_1, \vx_2$ and $\vx_3,\vx_4$ will only be very weakly correlated for large $r$ separation of the pairs of points, so the third term in Eq.~\eqref{expapprox} can be neglected.



This allows us to find the following estimate for the lensed $\tau_{NL}$ trispectrum,
\begin{multline}
\tilde{p}_{l_1 l_2}^{l_3 l_4}(L)\approx\int \frac{d^2 \br_{21}d^2 \br_{43} d^2\bl_{21}' d^2\bl_{43}'  }{(2\pi)^4} p_{l_1' l_2'}^{l_3' l_4'}(L)
e^{i\br_{21}.(\bl_{21}'-\bl_{21})}e^{i\br_{43}.(\bl_{43}'-\bl_{43})}\\
\times
\exp\Big(-\frac{1}{2}\langle (\bl_{21}'.\valpha_{21})^2\rangle\Big) \exp\Big(-\frac{1}{2}\langle(\bl_{43}'.\valpha_{43})^2\rangle\Big),
\end{multline}
where the expectation values can be calculated exactly as for power spectrum and bispectrum lensing. Since the trispectrum is well approximated
as involving only the CMB power spectrum on the small-scale modes (see Appendix~\ref{appendix:taunl}), this is consistent with lensing simply replacing the unlensed power spectra by the lensed spectra.


\subsection{$g_{NL}$ trispectrum}
The $\gnl$ trispectrum can be generated by local cubic corrections to a Gaussian field. It essentially measures the correlation of a small-scale local $\fnl$ with the large-scale field. Since local $\fnl$ itself has all the signal in squeezed shapes, this means that the signal in $\gnl$ is dominated by shapes with $l_1, l_2\ll l_3,l_4$, though unlike in the $\taunl$ trispectrum the signal is not all in $\l_1\sim 1$. For simplicity we will use the flat-sky approximation here.

We can approximate the primordial $\gnl$ signal by taking the large-scale curvature modes with wavenumbers $\vl_1$ and $\vl_2$ to be constant through last scattering.
For a primordial curvature perturbation $\zeta = \zeta_0 + \gnl \zeta_0^3$ the contribution to the trispectrum in this regime is then determined by
\bea
\la T(\vl_1)T(\vl_2)T(\vl_3)T(\vl_4)\ra_c &\approx&
C_{l_1}^{T\zeta_*} C_{l_2}^{T\zeta_*} \left\la \frac{\delta}{\delta\zeta_*(\vl_2)^*} \frac{\delta}{\delta\zeta_*(\vl_1)^*}\left(
T(\vl_3)T(\vl_4)\right)\right\ra \nonumber \\
&\approx&
\frac{6}{(2\pi)^2}g_{NL} C_{l_1}^{T\zeta_*} C_{l_2}^{T\zeta_*} (C_{l_3}+C_{l_4})\delta(\vl_1+\vl_2+\vl_3+\vl_4),
\label{gnlapprox}
\eea
where $\zeta_*$ is the primordial curvature perturbation at last scattering. This is a good approximation for $l_1,l_2 \ll 100$, but breaks down for the contributions where $l_2\agt 100$ where the finite thickness of last-scattering becomes important. However it is sufficient for ballpark estimates of the signal to noise and bias.

\subsubsection{Lensing bias}

With $l_1,l_2 \le l_3,l_4$, the $\gnl$-type trispectrum due to lensing is of the form
\be
\langle \Ltemp(\bl_1)\Ltemp(\bl_2)\Ltemp(\bl_3)\Ltemp(\bl_4)\rangle_c \approx
C_{l_1}^{\temp\psi} C_{l_2}^{\temp\psi} \left\la\frac{\delta }{\delta \psi(\vl_1)^* \delta\psi(\vl_2)^*} \left( \Ltemp(\vl_3)\Ltemp(\vl_4)\right)   \right\ra,
\ee
where $\psi$ is the lensing potential that gives the deflection angle $\valpha = \vgrad \psi$. Using
\be
\frac{\delta}{\delta \psi(\vl)^*} e^{i\cdot \vl' \cdot (\vx + \valpha)} = \frac{1}{2\pi}\vl\cdot \vl' e^{i\vl' \cdot (\vx + \valpha)} e^{-i\vl\cdot\vx}
\ee
and evaluating to leading order we have
\begin{multline}
\langle \Ltemp(\bl_1)\Ltemp(\bl_2)\Ltemp(\bl_3)\Ltemp(\bl_4)\rangle_c \approx
\frac{1}{(2\pi)^2} C_{l_1}^{\temp\psi} C_{l_2}^{\temp\psi}\biggl[ \vl_1\cdot \vl_4 \,\,\vl_2\cdot \vl_4 {C}_{l_4} + \vl_1\cdot \vl_3 \,\,\vl_2\cdot \vl_3 {C}_{l_3}
\\
- \vl_1\cdot(\vl_1+\vl_3)\vl_2\cdot(\vl_1+\vl_3) C_{|\vl_1+\vl_3|}
- \vl_1\cdot(\vl_1+\vl_4)\vl_2\cdot(\vl_1+\vl_4) C_{|\vl_1+\vl_4|}
\biggr]
\delta(\vl_1+\vl_2+\vl_3+\vl_4).
\label{gnl_lensing_shape}
\end{multline}
Note that for $l_1,l_2\ll l_3,l_4$ the two individually large pairs of terms nearly cancel. This leaves the total second-order response of the small-scale power being small, as expected. Numerically the Fisher signal to noise on this part of the lensing trispectrum for $\lmax=2000$ gives $S/N \ll 1$, and hence though much larger than likely primordial signals, is guaranteed to project onto $\gnl$ at much less than the $\gnl$ error bar (numerically we find a bias $\la \hat{g}_{\rm NL}\ra \sim \clo(100)$ where the error bar is $\sigma_{\gnl} \sim \clo(10^5)$ \cite{Fergusson:2010gn}).

In addition to Eq.~\eqref{gnl_lensing_shape} the lensing trispectrum~\cite{Zaldarriaga:2000ud}
\bea
p_{l_1 l_2}^{l_2 l_4}(L)&\approx&
-C_L^{\psi}(\vl_1\cdot \vL \tC_{l_1} + \vl_2\cdot \vL \tC_{l_2})(\vl_3\cdot \vL \tC_{l_3} + \vl_4\cdot \vL \tC_{l_4}) +\text{perms}
\eea
that contaminated $\taunl$ will also contaminate $\gnl$ for smaller $l_1,l_2$. For the configurations that contribute to $\gnl$ at low $l_1,l_2$ the signal-to-noise in the lensing is however also below unity, and hence the projection onto $\gnl$ is also guaranteed to be much smaller than the error bar.

We conclude that lensing bias on $\gnl$, though much larger than likely primordial signals, is small enough compared to the error bar to be neglected.

\subsubsection{Lensing of $\gnl$}

For $l_1 \alt 500$ we expect the unlensed short-leg approximation to be very accurate, and we now make this approximation in both of the $\gnl$ short legs labelled by $l_1$ and $l_2$, with index ordering convention $l_1,l_2 \le l_3,l_4$:
\begin{multline}
\langle \Ltemp(\bl_1)\Ltemp(\bl_2)\Ltemp(\bl_3)\Ltemp(\bl_4)\rangle_c \approx
\langle \temp(\bl_1)\temp(\bl_2)\Ltemp(\bl_3)\Ltemp(\bl_4)\rangle_c
\\=
\int \frac{ d^2 \bx_3 d^2 \bx_4 d^2 \bl_3' d^2 \bl_4'}{(2\pi)^6}\delta(\bl_1 +\bl_2+\bl_3'+\bl_4')t_{l_1 l_2 l_3' l_4'} e^{i(\bl_3'-\bl_3).\bx_3}e^{i(\bl_4'-\bl_4).\bx_4} \langle  e^{i\bl_3'.\valpha_3} e^{i\bl_4'.\valpha_4}\rangle.
\end{multline}
The average effect of small-scale lensing is therefore very much like the squeezed bispectrum, with just the two short-scale modes being lensed, consistent with replacing unlensed with lensed power spectra in Eq.~\eqref{gnlapprox}. An approximation for the lensed trispectrum therefore follows from corresponding approximations:
\be
\tilde{t}_{l_1 l_2 l_3 l_4} \approx \int \frac{\ud^2 \vr}{2\pi} \frac{\ud^2 \vl_{43}'}{2\pi} t_{l_1 l_2 l_3' l_4'}
\exp\left(-\frac{1}{2}\la (\vl'_{43}\cdot \valpha_{43})^2\ra\right),
\ee
which if desired could be evaluated in exactly the same way as for the bispectrum. However since observational constraints on $\gnl$ are very weak, and we know that local $\fnl$ estimates are not much affected by lensing on average, modelling the effect in detail is probably unnecessary.

\section{Conclusion}

We have discussed two effects of lensing. Firstly, the presence of large-scale lensing modes $\kappa_{LM}$ modulate the small-scale CMB and hence partially mimic the effect of primordial modulations,  giving rise to trispectrum and bispectrum signal that is potentially a source of confusion with primordial non-Gaussianity. The lensing effect is well-known to project onto the local non-Gaussianity model at the significant level of $\fnl\sim 9$, and we have now also calculated the projection onto $\taunl$; the signal is large compared to likely levels of primordial signal, however since the cosmic variance errors are large, the bias is only $\sim 10\%$ of the size of the error bar for Planck, and hence relatively not as important as for the bispectrum. For $\gnl$ the effect is probably negligible. Secondly, the effect of many smaller-scale lensing modes changes the detailed shape of any primordial bispectrum and trispectrum, similar to the smoothing effect on the power spectrum. For squeezed shapes we showed how this effect can easily be calculated, and that it is well approximated as applying CMB power spectrum lensing to each slice of the $m=0$ isotropic part of the squeezed bispectrum.

The potential bias on primordial signals due to the average change in the primordial shape due to CMB lensing is however very small. This is easily understood: since CMB lensing just moves points around, the $n$th moment of the temperature at any given point $\la [T(\vx)]^n\ra$ is unchanged under lensing on average, for example the total power is conserved:
\be
\la T(\vx)^2\ra = \int \ud \ln l\, \frac{l^2 C_l}{2\pi} = \int \ud \ln l\, \frac{l^2 \tilde{C}_l}{2\pi}.
\ee
 Local non-Gaussianity looks like a modulation of small-scale power as a function of position. In the squeezed limit we are considering very large-scale modulations, and the simplest way to measure the modulation is just to calculate the small-scale fluctuation variance as a function of position. However this variance is unchanged under lensing, and hence the non-Gaussianity estimate is unchanged under lensing on average, so it is unbiased. The total local skewness  $\la [T(\vx)]^3\ra$ is also invariant under lensing on average, giving\footnote{This is consistent with the exact bispectrum lensing result of Eq.~\eqref{exact_bispectrum} because $\sigma^2(0)=\Cgltwo(0)=0$. Note also that only the isotropic part of the bispectrum contributes to the skewness.}
\be
\la T(\vx)^3\ra = \int \frac{\ud \vl_1^2  \ud \vl^2}{(2\pi)^4} \, b_{l_1 l_2 l_3} =
\int \frac{\ud \vl_1^2  \ud \vl^2}{(2\pi)^4} \, \tilde{b}_{l_1 l_2 l_3}.
\ee
In reality the picture is a bit more complicated because estimators weight by the inverse signal plus noise, so the estimators are not exactly unbiased, but nonetheless there is a good reason why the bias is expected to be small for local shapes. The detailed change in shape due to lensing could in principle be detected if $\fnl \agt 20$, but the change is almost orthogonal to the unlensed shape and hence neglecting it is usually harmless. On the other hand the bias due to correlation between primordial and lensing-induced non-Gaussianity should be subtracted or consistently modelled to avoid biases, especially for the bispectrum and the $L=1$ dipole-modulation part of the $\taunl$ trispectrum.

Lensing also affects the variance of non-Gaussianity estimators; this has been calculated in Ref.~\cite{Hanson:2009kg}, and optimized bispectrum estimators accounting for this extra variance have been derived in Ref.~\cite{Lewis:2011fk}.

\section{Acknowledgements}
AL thanks Duncan Hanson and Anthony Challinor for discussion, and for sharing the numerical results of Ref.~\cite{Hanson:2009kg} shown in Fig.~\ref{perturbative_compare}.
Some of the results in this paper have been derived using HealPix~\cite{Gorski:2004by}.
AL and DR acknowledge support from the Science and Technology Facilities Council [grant number ST/I000976/1], and RP via a research studentship.
Some of the calculations for paper  were performed on the
 COSMOS Consortium supercomputer within the DiRAC Facility jointly
 funded by STFC, the Large Facilities Capital Fund of BIS and the
 University of Sussex.

\appendix

\section{General lensed temperature bispectrum}
\label{any_shape}
For general shapes, all three temperature multipoles need to be lensed, giving
\begin{multline}
\la \Ltemp(\vl_1) \Ltemp(\vl_2) \Ltemp(\vl_3) \ra=
\int \frac{d^2\vx_1 }{2\pi} \frac{d^2\vx_2 }{2\pi}\frac{d^2\vx_3 }{2\pi}
\frac{d^2\vl_1' }{2\pi}\frac{d^2\vl_2' }{2\pi}\frac{d^2\vl_3' }{2\pi} \times \\
\la \temp(\vl_1') \temp(\vl_2') \temp(\vl_3') e^{-i\vl_1\cdot \vx_1}e^{-i\vl_2\cdot \vx_2}e^{-i\vl_3\cdot \vx_3}
e^{i\vl_1'\cdot (\vx_1+\valpha_1)} e^{i\vl_2'\cdot (\vx_2+\valpha_2)} e^{i\vl_3'\cdot (\vx_3+\valpha_3)} \ra.
\end{multline}
As usual we take $\valpha$ to be uncorrelated to $\temp$ to isolate the effect of lensing on the primordial bispectrum.

Let's define $\vl'\equiv (\vl_2'-\vl_3')/2 = \vl_2'+\vl_1'/2$,
$\va_2 \equiv \valpha_2 -\valpha_1$, $\va_3\equiv \valpha_3-\valpha_1$. Then
\be
\vl_1'\cdot \valpha_1 + \vl_2'\cdot \valpha_2 + \vl_3'\cdot \valpha_3 =
\vl'\cdot(\va_2-\va_3) -\frac{\vl_1'}{2}\cdot(\va_2+\va_3),
\ee
\be
\vx_1\cdot(\vl_1-\vl_1') + \vx_2\cdot(\vl_2-\vl_2') +\vx_3\cdot(\vl_3-\vl_3')
=\left( \vx_1 - \frac{\vx_2+\vx_3}{2}\right)\cdot(\vl_1-\vl_1') + (\vx_2-\vx_3)\cdot (\vl-\vl') ,
\ee
and hence defining $\vr\equiv -\left( \vx_1 - \frac{\vx_2+\vx_3}{2}\right)=  \frac{(\vr_{31}+\vr_{21})}{2}$
\ba
\lb_{l_1 l_2 l_3} \!&=&
\!\int \frac{d^2\vr_{32} }{2\pi} \frac{d^2\vr }{2\pi}  \frac{d^2\vl_1' }{2\pi} \frac{d^2\vl' }{2\pi}
 b_{l_1' l_2' l_3'}
 e^{i \vr_{32}\cdot(\vl'-\vl)}e^{i \vr \cdot(\vl_1'-\vl_1)}
    \exp\left( - \frac{1}{2}\left\la\left[ \vl'\cdot(\va_2-\va_3) - \frac{\vl_1'}{2}\cdot(\va_2+\va_3)\right]^2\right\ra\right)\nonumber \\
 &=&
\! \int \frac{d^2\vr_{32}  }{2\pi} \frac{d^2\vr }{2\pi} \frac{d^2\vl' }{2\pi}\frac{d^2\vl_1' }{2\pi}
 b_{l_1' l_2' l_3'}
 e^{i \vr_{32}\cdot(\vl'-\vl)}
e^{i \vr \cdot(\vl_1'-\vl_1)}
  \exp\left( - \frac{1}{2}\left[
 l'{}^2\left(\sigma^2(r_{32}) + \cos 2\phi_{l'r_{32}} \Cgltwo(r_{32})\right)
 \right]\right)\times
 \nonumber\\
 &&
\exp\left( - \frac{l_1'{}^2}{4}\biggl[
  \sigma^2(r_{21}) + \cos 2\phi_{l_1'r_{21}} \Cgltwo(r_{21})
 + \sigma^2(r_{31}) + \cos 2\phi_{l_1'r_{31}} \Cgltwo(r_{31})  \right.
\nonumber \\
 &&\hspace{8cm}
 \left.
 -\frac{1}{2}\left\{\sigma^2(r_{32}) + \cos 2\phi_{l_1'r_{32}} \Cgltwo(r_{32})\right\}
 \biggr]\right).
\label{exact_bispectrum}
\ea
This gives the general result of lensing of the bispectrum. For squeezed bispectra, to $\clo(l_1^2/l^2)$ we can easily recover the result Eq.~\eqref{blens_squeezed} from the unlensed short-leg approximation since
\ba
\lb_{l_1 l_2 l_3} \!&\approx&
\! \int \frac{d^2\vr_{32}  }{2\pi} \frac{d^2\vr }{2\pi} \frac{d^2\vl' }{2\pi}\frac{d^2\vl_1' }{2\pi}
 b_{l_1' l_2' l_3'}
 e^{i \vr_{32}\cdot(\vl'-\vl)}
e^{i \vr \cdot(\vl_1'-\vl_1)}
  \exp\left( - \frac{1}{2}\left[
 l'{}^2\left(\sigma^2(r_{32}) + \cos 2\phi_{l'r_{32}} \Cgltwo(r_{32})\right)
 \right]\right)
 \nonumber\\
 &=&
 \! \int \frac{d^2\vr_{32}  }{2\pi} \frac{d^2\vl' }{2\pi}
 b_{l_1 l_2' l_3'}
 e^{i \vr_{32}\cdot(\vl'-\vl)}
  \exp\left( - \frac{1}{2}\left[
 l'{}^2\left(\sigma^2(r_{32}) + \cos 2\phi_{l'r_{32}} \Cgltwo(r_{32})\right)
 \right]\right)
\ea

\section{Angular decomposition of squeezed triangles on the full sky}
\label{curved_sky}

On the flat sky a bispectrum triangle can be parameterized in terms of the wavenumbers of the large-scale and short-scale modes, and the angle between them. On the full sky we can change from $l_1,l_2,l_3$ to $l_1, L\equiv (l_2+l_3)/2$, and $M\equiv l_3-l_2$, where $|M|\le l_1$. For $l_1$ even, $L$ is an integer, corresponding to being able to divide the long wavenumber in two, and $M$ is even; for odd $l_1$, $L$ is half-integer and $M$ is only odd. On the full sky the reduced bispectrum is defined by
\be
b(l_1,L,M) = b_{l_1 l_2 l_3} = \sqrt{\frac{4\pi}{(2l_1+1)(2l_2+1)(2l_3+1)}}\threejz{l_1}{l_2}{l_3}^{-1}B_{l_1 l_2 l_3},
\ee
where we note that this change of variable is related to the Regge symmetries~\cite{Bincer1970,AngularMom} of the $3j$ symbols since
\ba
\threejz{l_1}{l_2}{l_3} = \threej{l_1}{L}{L}{M}{-M/2}{-M/2}.
\ea
Using  $l_1,L,M$ we can replace sums over wavenumbers using
\be
\sum_{l_1 l_2 l_3} = \sum_{l_1}\sum_{L} \sum_{M=-l_1}^{l_1}
\ee
and
\be
(2l_1+1)(2l_2+1)(2l_3+1) = (2l_1+1)[(2L+1)^2- M^2].
\ee
For squeezed triangles $l_1 \ll L$, $|M|\ll L$, and to $\clo(l_1^2/L^2)$ corrections in the flat sky limit\footnote{More generally one can define $\cos \phi \equiv 2M/(2l_1+1)$; as one approaches the flat-sky limit the angular dependence of the 3j symbol ensures that $\sum_M \threejz{l_1}{l_2}{l_3}^2[\dots] \propto \int \ud\phi [\dots]$; i.e. roughly that $\threejz{l_1}{l_2}{l_3}^2 \propto 1/\sin(\phi)$ so that the weight per $M$ mode is much larger for $\sin\phi \sim 0$ (i.e. $|M|\sim l_1$) than for $M\sim 0$, reflecting the fact that there are more triangles with integer sides for near-isosceles than near-flat squeezed shapes at fixed $L$.}  $M = l_1 \cos(\phi_{ll_1})$.

%
The Fisher matrix gives the expected error on a bispectrum estimate, and the overlap function $F(B,B')$ quantifies the bias when a bispectrum $B$ is estimated and a different bispectrum shape $B'$ is present.
For squeezed triangles (with $l_1 \ll 200$, corresponding to the spacing of the acoustic peaks), $C_{l_2} C_{l_3} = C_L^2 +\clo(l_1^2/L^2)$, so the overlap function is given by \be
F(B,B') =\frac{1}{6}\sum_{l_1 l_2 l_3} \frac{B_{l_1 l_2 l_3} B_{l_1 l_2 l_3}'}{ C_{l_1} C_{l_2} C_{l_3}}\approx
\frac{1}{6} \sum_{l_1 L}  \frac{(2l_1+1) (2L+1)^2}{ C_{l_1} C_{L}^2} \sum_M \threej{l_1}{L}{L}{M}{-M/2}{-M/2}^2 b(l_1,L,M)  b(l_1,L,M) '.
\label{overlap_fullsky}
\ee
The $3j$ term factorizes into terms involving $M$ only through a combination of $l_1$ and $M$
\be
\threej{l_1}{l_2}{l_3}{0}{0}{0}^2= \threej{l_1}{L}{L}{M}{-M/2}{-M/2}^2
= \frac{(2L-l_1)!}{(2L+l_1+1)!}\left(\frac{ ([L+l_1]/2)!}{([L-l_1]/2)!}\right)^2 \left(\frac{(l_1-M)!(l_1+M)!}{\left(([l_1-M]/2)!([l_1+M]/2)! \right)^2} \right),
\ee
so we can define polynomials $P_m(M)$ in $M$ that depend only on $M$ and $l_1$ such that
\ba
\sum_{M} \threej{l_1}{L}{L}{M}{-M/2}{-M/2}^2  P_m(M) P_{m'}(M) \propto \delta_{m m'}
\ea
for $i,j \le l_1$. Expanding the bispectrum in terms of $P_m$, this relation ensures that bispectra with different $m$ components are orthogonal to $\clo(l_1^2/L^2)$ (Eq.~\eqref{overlap_fullsky} gives zero). Explicitly we can define $P_0\equiv 1$ and
\ba
l_1^2 P_2(M) \equiv 2 M^2 - l_1(l_1+1)
\qquad
l_1^4 P_4(M) \equiv 8 M^4 - 8(l_1-1)(l_1+2)M^2 +  l_1(l_1+1)(l_1+3)(l_1-2), \qquad \dots
\ea
where the $l_1$ factors on the LHS are for convenience so that in the flat sky limit $P_m(M)= \cos (m\phi)$. Thus we define the angular moments of the bispectrum on the full sky by
\ba
b^m_{l_1 L} \equiv  \frac{\sum_M \threej{l_1}{L}{L}{M}{-M/2}{-M/2}^2 P_m(M) b_{l_1 l_2 l_3} }
{  \sum_M \threej{l_1}{L}{L}{M}{-M/2}{-M/2}^2 P_m(M)^2 }
\ea
(if desired the normalization can be calculated analytically).


As an example of an explicitly anisotropic bispectrum, the full-sky temperature lensing reduced bispectrum can be expanded in the squeezed limit (where most of the signal is) as
\ba
b_{l_1 l_2 l_3} &=& \frac{1}{2}[ l_1(l_1+1) + l_2(l_2+1)-l_3(l_3+1)] C_{l_1}^{T\psi} \tilde{C}_{l_2} + \text{perms}.
\\
&\approx& C_{l_1}^{T\psi} \frac{1}{2}\left[ l_1(l_1+1)(\tilde{C}_{l_2}+\tilde{C}_{l_3})   - M(2L+1)(\tilde{C}_{l_2}-\tilde{C}_{l_3})   \right]\\
&\approx& C_{l_1}^{T\psi} \left( l_1(l_1+1) \tilde{C}_L +  \frac{1}{2}M^2(2L+1) \frac{d \tilde{C}_L}{d L}\right) \\
&\approx&
\frac{1}{2}l_1(l_1+1) C_{l_1}^{T\psi} \left(
\left[\frac{2 M^2}{l_1(l_1+1)} - 1\right] \frac{ \ud \tilde{C}_L}{\ud \ln (2L+1)}
+  \frac{1}{L(L+1)} \frac{ \ud[L(L+1)\tilde{C}_L]}{\ud\ln(2L+1)}
\right),
\ea
and hence has $P_0(M)$ and $P_2(M)$ components, corresponding to the scalar and quadrupolar angular dependence expected from magnification and shear. Local non-Gaussianity also gives significant $|m|\ge 2$ components on scales where $l_1$ is sub-horizon at recombination (see Fig.~\ref{moment_slices}).

\section{Squeezed approximation for $\taunl$ CMB trispectrum}
\label{appendix:taunl}

Here we approximate the full expression for $\taunl$ and relate to the modulation estimator (c.f. Ref.~\cite{Okamoto:2002ik}). The unlensed $\taunl$ reduced trispectrum is~\cite{Okamoto:2002ik,Regan:2010cn}
\be \label{eq:taunltrisp}
p_{l_3l_4}^{l_1 l_2}(L) = \taunl \int \ud r_1 \ud r_2 r_1^2 r_2^2
  F_L(r_1,r_2)\left[\alpha_{l_1}(r_1)\beta_{l_2}(r_1)+\beta_{l_1}(r_1)\alpha_{l_2}(r_1)\right]\left[ \alpha_{l_3}(r_2)\beta_{l_4}(r_2)
+\beta_{l_3}(r_2)\alpha_{l_4}(r_2)\right]
\ee
where
\begin{eqnarray}
\alpha_{l}(r) &\equiv& 4\pi \int \ud \ln k j_l(kr) \frac{k^3\Delta_{l}(k)}{2\pi^2}\nonumber \\
\beta_{l}(r) &\equiv& 4\pi \int \ud \ln k j_l(kr) \Delta_{l}(k) \clp_{\zeta}(k)
\label{alphabetadef}
\end{eqnarray}
and
\be
F_L(r_1,r_2) \equiv 4\pi \int \ud \ln k \clp_{\zeta}(k) j_L(kr_1) j_L(k r_2).
\ee
We are interested in very low $L$ where all the signal to noise is, so that $L\ll l_1,l_2,l_3, l_4$. For small scales $\alpha_l(r)$ and $\beta_l(r)$ are sharply peaked at recombination, but $F_L(r_1,r_2)$ varies only very slowly on the scale of the large-scale modulation modes.
Therefore
\be
p_{l_3l_4}^{l_1 l_2}(L) \approx \taunl F_L(r_*,r_*)\int \ud r_1 r_1^2
    [\alpha_{l_1}(r_1) \beta_{l_2}(r_1)+\beta_{l_1}(r_1) \alpha_{l_2}(r_1)] \int \ud r_2 r_2^2 \left[\alpha_{l_3}(r_2)\beta_{l_4}(r_2)+\beta_{l_3}(r_2)\alpha_{l_4}(r_2)\right].
\ee
Note that $F_L(r_*,r_*) = C_L^{\zeta_*}$, i.e. the angular power spectrum of the primordial curvature perturbations at recombination. Also since $L$ is very small, much smaller than the scale of variation of the $C_l$, $l_1\approx l_2$ and so we have
\be
\int r^2 \ud r \left[\alpha_{l_1}(r)\beta_{l_2}(r)+\beta_{l_1}(r)\alpha_{l_2}(r)\right]\sim C_{l_1} +C_{l_2}.
\label{alphabetaint}
\ee
This is exact for $l_1=l_2$, and accurate to sub-percent level for $|l_1-l_2|<10$ that is relevant for $\taunl$.
Hence
\be
p_{l_3l_4}^{l_1 l_2}(L) \approx \taunl F_L(r_*,r_*) (C_{l_1}+C_{l_2})(C_{l_3}+ C_{l_4})\approx \taunl C_L^{\zeta_*} (C_{l_1}+C_{l_2})(C_{l_3}+ C_{l_4}).
\ee
This agrees with the approximation and definition of Eqs.~\eqref{taunl_L}~\eqref{Tmodulation} in the main text.
 For $L=1$ the relevant scales controlling the approximations are the thickness of the last-scattering surface ($\sim 100\Mpc$) and the distance to the last-scattering surface $r_*\sim 14 000\Mpc$, and hence approximations are expected to be accurate at the percent level for high $l_1,l_2,l_3,l_4$ ($\gg 100$). The approximate effect of lensing is also clear: the small-scale temperature modes contribute to the trispectrum via their power spectrum, and averaging over small-scale lensing modes simply changes the unlensed power spectra to the lensed power spectra.

\providecommand{\aj}{Astron. J. }\providecommand{\apj}{Astrophys. J.
  }\providecommand{\apjl}{Astrophys. J.
  }\providecommand{\mnras}{MNRAS}\providecommand{\aap}{Astron. Astrophys.}

\end{document}

%% file: macros.tex
\newcommand{\Mpc}{\text{Mpc}}
\newcommand{\half}{{\textstyle \frac{1}{2}}}

\newcommand{\lmax}{l_{\text{max}}}

\newcommand{\fnl}{f_{\rm{NL}}}
\newcommand{\taunl}{\tau_{\rm{NL}}}
\newcommand{\gnl}{g_{\rm{NL}}}

\providecommand{\CAMB}{\textsc{camb}}

\newcommand{\begm}{\begin{pmatrix}}
\newcommand{\enm}{\end{pmatrix}}
\newcommand{\threej}[6]{{\begm #1 & #2 & #3 \\ #4 & #5 & #6 \enm}}
\newcommand{\threejz}[3]{{\begm #1 & #2 & #3 \\ 0 & 0 & 0 \enm}}

\newcommand\ba{\begin{eqnarray}}
\newcommand\ea{\end{eqnarray}}
\newcommand\bea{\begin{eqnarray}}
\newcommand\eea{\end{eqnarray}}

\newcommand\be{\begin{equation}}
\newcommand\ee{\end{equation}}

\newcommand{\valpha}{{\boldsymbol{\alpha}}}
\newcommand{\vgrad}{{\boldsymbol{\nabla}}}

\providecommand{\var}{\text{var}}
\providecommand{\cov}{\text{cov}}

\newcommand{\la}{\langle}
\newcommand{\ra}{\rangle}

\newcommand{\ud}{{\rm d}}

\newcommand{\boldvec}[1]{{{\mathbf{#1}}}}

\newcommand{\vL}{\boldvec{L}}

\newcommand{\va}{\boldvec{a}}

\newcommand{\vh}{\boldvec{h}}

\newcommand{\vk}{\boldvec{k}}
\newcommand{\vl}{\boldvec{l}}

\newcommand{\vn}{\boldvec{n}}

\renewcommand{\vr}{\boldvec{r}}

\newcommand{\vx}{\boldvec{x}}

\newcommand{\clf}{\mathcal{F}}

\newcommand{\clk}{\mathcal{K}}

\newcommand{\clo}{\mathcal{O}}
\newcommand{\clp}{\mathcal{P}}

\newcommand{\vnhat}{\hat{\vn}}